\documentclass[aps,prd,reprint,twocolumn,preprintnumbers,floatfix,nofootinbib]{revtex4-1}
\usepackage[utf8]{inputenc}
\usepackage[dvips]{graphicx}
\usepackage[dvipsnames]{xcolor}
\usepackage{color}
\usepackage{relsize}
\usepackage{graphics}
\usepackage{epstopdf}
\usepackage{hyperref}
\usepackage{mathrsfs}
\usepackage{amssymb}
\usepackage{physics}
\usepackage{booktabs}

\usepackage[normalem]{ ulem }
\usepackage{amsthm}
\usepackage{amsmath}
\usepackage{cancel}
\usepackage{fontawesome} 
\usepackage[caption=false]{subfig}
\usepackage{tabularx,ragged2e} 
 \newcolumntype{L}{>{\RaggedRight\arraybackslash}X}

\usepackage{hyperref}

\newcommand{\be}{\begin{equation}}
\newcommand{\ee}{\end{equation}}
\newcommand{\bea}{\begin{eqnarray}}
\newcommand{\eea}{\end{eqnarray}}
\newcommand{\ba}{\begin{aligned}}
\newcommand{\ea}{\end{aligned}}

\newcommand{\TBH}{T_{\rm BH}}
\newcommand{\MBH}{M}
\newcommand{\MBHi}{M_{\rm in}}
\newcommand{\TBHi}{T_{\rm BH}^{\rm in}}

\newcommand{\MPL}{M_p}

\newcommand{\as}{a_\star}
\newcommand{\asi}{a_\star^{\rm in}}

\newcommand{\fBH}{f_\mathrm{PBH}}
\newcommand{\Neff}{N_{\rm eff}}

\newcommand{\figref}[1]{Fig.~\ref{#1}}

\newcommand{\pd}[2]{\frac{\partial #1}{\partial #2}}
\newcommand{\fd}[2]{\frac{d #1}{d #2}}
\newcommand{\secref}[1]{Sec.~\ref{#1}}

\newcommand\myshade{80}
\colorlet{mylinkcolor}{ForestGreen}
\colorlet{mycitecolor}{Red}
\colorlet{myurlcolor}{violet}

\hypersetup{
  linkcolor  = mylinkcolor!\myshade!black,
  citecolor  = mycitecolor!\myshade!black,
  urlcolor   = myurlcolor!\myshade!black,
  colorlinks = true
}

\usepackage{pgfplots}
\newcommand{\equaref}[1]{Eq.~(\ref{#1})}

\pgfplotsset{compat=1.17}

\begin{document}
\sloppy  

\preprint{IPPP/22/79}

\vspace*{1mm}

\title{
Evaporation of Primordial Black Holes  in the Early Universe: \\Mass and Spin Distributions 
}

\author{Andrew Cheek$^{a}$}
\email{acheek@camk.edu.pl}
\author{Lucien Heurtier$^{b}$}
\email{lucien.heurtier@durham.ac.uk}
\author{Yuber F. Perez-Gonzalez$^{b}$}
\email{yuber.f.perez-gonzalez@durham.ac.uk}
\author{Jessica Turner$^{b}$}
\email{jessica.turner@durham.ac.uk}

\affiliation{$^a$ Astrocent, Nicolaus Copernicus Astronomical Center of the Polish Academy of Sciences, ul.Rektorska 4, 00-614 Warsaw, Poland}
\affiliation{$^b$ Institute for Particle Physics Phenomenology, Durham University, South Road DH1 3LE, Durham, U.K.}

\begin{abstract}
{Many cosmological phenomena lead to the production of primordial black holes in the early Universe. These phenomena often create a population of black holes with extended mass and spin distributions. As these black holes evaporate via Hawking radiation, they can modify various cosmological observables, lead to the production of dark matter, modify the number of effective relativistic degrees of freedom, $\Neff$, source a stochastic gravitational wave background and alter the dynamics of baryogenesis. We consider the Hawking evaporation of primordial black holes that feature non-trivial \emph{mass and spin} distributions in the early Universe. We demonstrate that the shape of such a distribution can strongly affect most of the aforementioned cosmological observables. We outline the numerical machinery we use to undertake this task. We also release a new version of {\tt FRISBHEE} \href{https://github.com/yfperezg/frisbhee}{\faGithub} that handles the evaporation of primordial black holes with an arbitrary \emph{mass and spin} distribution throughout cosmic history.
}
\end{abstract}

\maketitle

\section{Introduction}

The birth of gravitational wave astronomy~\cite{LIGOScientific:2016aoc, LIGOScientific:2017vwq} has produced a flurry of interest in Primordial Black Holes (PBHs)~\cite{Escriva:2022duf}. Unlike astrophysical black holes,  which result from stellar collapse, PBHs formed in the early Universe when large over-densities collapse under their Schwarzchild radius. If proven to exist, the implications for our understanding of the post-inflationary Universe would be tremendous. This is because the primordial power spectrum implied by the cosmic microwave background (CMB) is insufficient to produce them. Evidence of PBHs would represent a concrete way to move beyond the current model of inflationary cosmology~\cite{mambrini2021particles}. Measuring the PBH distribution would provide a direct view into their production, the early Universe~\cite{1982Natur.298..538C,Jedamzik:1996mr, carr:2019kxo} and inflation~\cite{Carr:1974nx,PhysRevD.50.7173,PhysRevD.48.543,Kawasaki:1997ju, Kawasaki:1998vx, Green:2000he,Bassett:2000ha, Allahverdi:2006iq, BuenoSanchez:2006rze,Kawasaki:2006zv, Jedamzik:2010dq, Easther:2010mr, Ezquiaga:2017fvi,Tada:2019amh,  Heurtier:2022rhf, Kristiano:2022maq}. Even after their initial production, BHs can accrete the Standard Model (SM) plasma, merge with their peers, acquiring rotational momentum~\cite{Flores:2021tmc,Hooper:2020evu}.  



PBHs can therefore feature various mass and spin distributions depending on the cosmological scenario considered. Whereas PBHs with masses larger than $10^{15}~\mathrm{g}$ are stable on cosmological time scales, lighter black holes may have evaporated via Hawking radiation before the current epoch~\cite{Hawking:1974rv,Hawking:1974sw}. The effect of this evaporation was studied in many different contexts (see e.g. Ref.~\cite{Auffinger:2022khh} for a recent review). PBHs with intermediate masses in the range $10^{8}-10^{12}\mathrm{g}$ are known to evaporate after Big-Bang Nucleosynthesis (BBN)~\cite{Keith:2020jww} and, therefore, cannot constitute a sizeable fraction of the Universe's energy density since, through Hawking radiation, they would change the neutron-to-proton ratio at the onset of BBN and therefore, the abundance of light elements that are measured today with excellent accuracy. However, a sizeable abundance of lighter PBHs may have formed in cosmic history without affecting the post-BBN era. Interestingly, such PBHs may even have dominated the energy density of the Universe before they evaporated, leading to a phase of early matter domination (EMD) and a subsequent reheating of the Universe. In both cases, several works have recently studied the imprints of such PBHs Hawking evaporation on particle and astrophysical  data by studying its effect on the dark matter (DM) relic density and phase space distribution~\cite{Cheek:2021odj, Cheek:2021cfe, Hooper:2019gtx, Masina:2020xhk,Morrison:2018xla, Auffinger:2020afu, Khlopov:2004tn, Allahverdi:2017sks, Lennon:2017tqq, Gondolo:2020uqv, Baldes:2020nuv, Bernal:2020bjf, Bernal:2020ili, Masina:2021zpu, Kitabayashi:2021hox, Bernal:2021bbv,Bernal:2020kse,Bernal:2022oha}, the effective number of relativistic degrees of freedom, $N_{\rm eff}$, at the time of CMB emission \cite{Cheek:2022dbx, Hooper:2019gtx, Hooper:2020evu, Masina:2020xhk, Arbey:2021ysg, Masina:2021zpu, Bhaumik:2022zdd,Cheek:2022dbx}, the dynamics of Baryo/Leptogenesis \cite{Perez-Gonzalez:2020vnz, JyotiDas:2021shi,Datta:2020bht, Morrison:2018xla, Fujita:2014hha, Granelli:2020pim, Hook:2014mla, Hamada:2016jnq,Hooper:2020otu,Chaudhuri:2020wjo,Bernal:2022pue,Barman:2022pdo,Gehrman:2022imk}, the hydrogen 21-cm line~\cite{Cang2021:2108.13256v2}, the production of gravitational waves~\cite{Inomata:2020lmk, Domenech:2021wkk, Bhaumik:2022pil}, and the electroweak vacuum stability~\cite{Dai:2019eei,Burda:2016mou,Hayashi:2020ocn}. In most of these studies, the distribution of PBHs considered was monochromatic, either in mass, or in spin, see, however, Refs.~\cite{Barrow:1991dn,Gutierrez:2017ibk} regarding the evaporation of a power-law mass distribution in the early Universe, Ref.~\cite{Mosbech:2022lfg} for a study of the impact of mass distributions on limits of PBH as DM, and Ref.~\cite{Boudon:2020qpo} for a first attempt in considering baryogenesis in the context of extended distributions.
In this paper, we consider the detailed evaporation of PBH populations that are not monochromatic, but instead spread over extended distributions in mass and/or in spin, tracking the evolution of the Universe carefully. 

The paper is organised as follows: In Sec.~\ref{sec:distributions} we review the most well-studied mass and spin distributions that have been reported in the literature, and that we will be considering throughout this work. We then review in \secref{sec:evaporation} the dynamics of the spin, $a_\star$, and mass, $M$, of a Kerr black hole during Hawking evaporation, and we extend this description in \secref{sec:Schwarz_BHs} to the evaporation of an extended distribution in the plane $(M,a_\star)$. Details regarding the  calculations are presented in \secref{sec:Schwarz_BHs} and expanded in Appendix  \ref{ap:KPBH_sol}, \ref{ap:jaco} and  \ref{ap:Nconservation},  which are used in our numerical simulations. In the remainder of the paper, we consider the various observables that can be affected by non-monochromatic PBH distributions. Specifically, in \secref{sec:cosmo} we explore the effect of extended mass distributions on the dynamics of the Universe when PBH that dominate the energy density evaporate and reheat the SM sector and follow in \secref{sec:DM} where we study similar effects on the production of DM and explore how the DM relic density is affected by the extension of the distribution in various examples. In particular, we explore how the constraints on warm DM from measurements of the Lyman-$\alpha$ forest are accordingly affected. In Sec.~\ref{sec:DR} we derive constraints on PBHs that have an extended distribution in mass and spin, by looking at their contribution to $\Delta N_{\rm eff}$. In \secref{sec:GWs} we comment on the possible imprints that extended distributions of PBHs could leave in the spectrum of primordial gravitational wave that could be observed in the future, and we finally conclude in \secref{sec:conc}.

Throughout, we use natural units where $\hbar = c = k_{\rm B} = 1$, and we define the Planck mass to be $M_p=1/\sqrt{G}$, with $G$ the gravitational constant.

\section{Extended PBH Distributions}
Depending on their formation mechanism, the PBHs may form with a mass~\cite{Carr:2020xqk} and spin distribution. In this section, we review some well-motivated PBH mass and spin distributions that we consider throughout this paper. Further, we assume that the mass and spin distributions form at the same time in the evolution of the Universe and their distributions can be convoluted. We will discuss these assumptions later on.

\subsection{Mass Distributions}\label{sec:distributions}
In what follows, we will, for each of the mass distributions considered, denote by $M_c$ the value of the mass at which the corresponding mass fraction $M\times\fBH(M)$ peaks, and by $\sigma$ its width in logarithmic space. 

\begin{figure*}[t!]
    \centering
    \includegraphics[width=\linewidth]{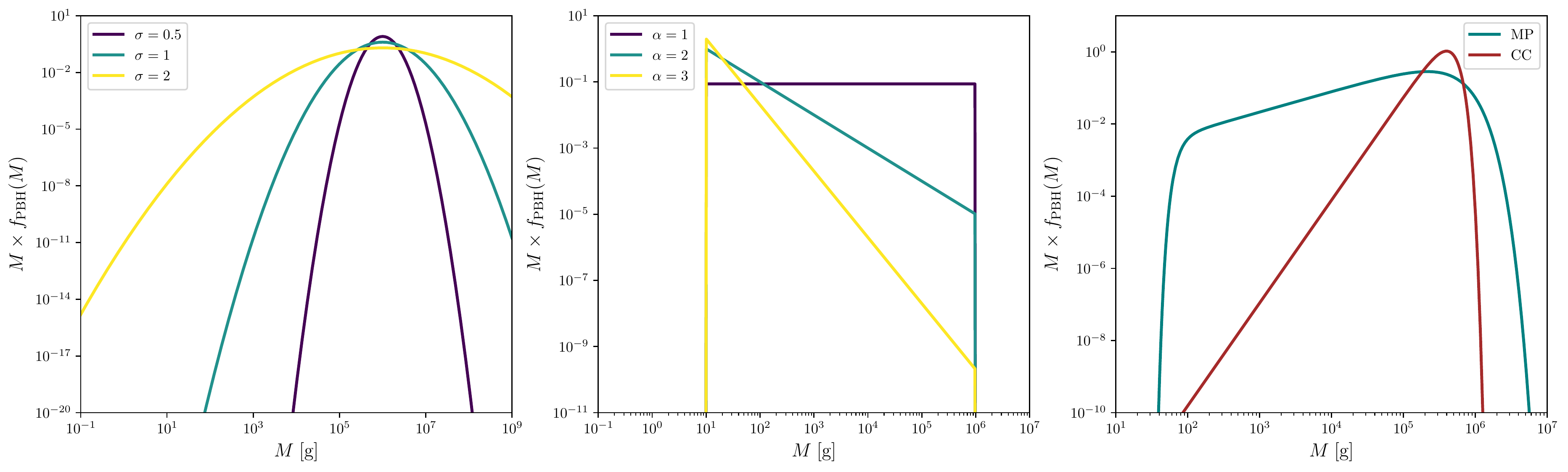}
    \caption{\label{fig:lognormal} \footnotesize The left plot shows the  log normal mass distribution with $M_c=10^{6}$ g for  $\sigma=0.5, 1.0, 2.0$ shown in violet, teal and yellow respectively. The central plot shows the power law mass distribution with $M_c=10$ g and $M_c 10^\sigma=10^6$ g for $\alpha=1, 2, 3$ shown in violet, teal and yellow, respectively. The right plot shows the mass distributions generated through metric preheating-\cite{Auclair:2020csm} (taken from Appendix C, Fig. 8) and critical collapse~\cite{Choptuik:1992jv,Yokoyama:1998qw,Yokoyama:1998xd}, using $M_c = 10^{5.6}~\mathrm{g}$ to align the two distributions.}
\end{figure*}

\bigskip
\paragraph{Log-Normal (LN)}~\newline~\newline
The production of PBHs from inflation usually requires the existence of a short period of {\em ultra-slow-roll} that produces a peak in the primordial power spectrum of scalar curvature perturbations~\cite{Ballesteros:2017fsr, Karam:2022nym, Dalianis:2018frf, Heurtier:2022rhf}. Generically, such peak is known to produce a  log-normal mass function \cite{Dolgov:1992pu} and this has been numerically and analytically verified for slow-roll inflation \cite{Green:2016xgy,Dolgov:2008wu}. 
The corresponding PBH mass distribution has the following form
\be\label{eq:distri_lognormal}
\fBH(M) = \frac{1}{\sqrt{2\pi}\sigma M}\exp\left[-\frac{\log^2(M/M_c)}{2\sigma^2}\right]\,,
\ee
where $M_c$ is the initial peak of the distribution and $\sigma$ is the width of the distribution. In the left most plot of \figref{fig:lognormal}, we show $M\times \fBH(M)$ as a function of $M$ for a central initial mass of PBH, $M_c=10^{6}$ g with varying values of the width $\sigma$.\\

\paragraph{Power-Law (PL)}~\newline~\newline
Another possible formation mechanism of PBHs is the case where a large scale-invariant power spectrum of primordial perturbations collapses in a Universe that is dominated by a perfect fluid with constant equation-of-state parameter, $w$. In that case, the distribution of PBHs takes the form \cite{Carr:1975qj}
\be\label{eq:distri_powerlaw}
\fBH(M)\propto 
\begin{cases}
M^{-\alpha}, & \text{for}~ M_c\leq M\leq M_c\times 10^{\sigma};\\
0, & \text{else,}
\end{cases}
\ee
where the exponent $\alpha$ is given by
\be
\alpha\equiv \frac{4w+2}{(w+1)}\,,
\ee
and the mass range $[M_c,\,M_c 10^\sigma]$ depends on the domain of frequencies over which this scale-invariant power spectrum was formed. Physical situations in which the Universe is not inflating anymore typically correspond to values of $w$ in the range $-1/3<w\leqslant 1$, and thus to a scaling exponent 
\be\label{eq:range}
1< \alpha \leqslant 3\,.
\ee
In the central plot of \figref{fig:lognormal}, we show the power law mass distribution for varying $\alpha$ values with $M_c (M_c 10^\sigma)$ taking the value $10 \, (10^6)$ g.  \\

\paragraph{Critical Collapse (CC)}~\newline~\newline
The application of critical scaling to gravitational collapse is thought to describe the process of PBH formation from primordial fluctuations in a rigorous way~\cite{Choptuik:1992jv,Yokoyama:1998qw,Yokoyama:1998xd, Kuhnel:2015vtw}. Traditionally, overdensities were assumed to produce PBHs that had the same mass as their horizon mass. Instead, there is an upper cut-off at around the horizon mass but then a tail in the distribution for lower masses, this is a fairly generic finding over many inflationary models~\cite{Kuhnel:2015vtw}. The resulting PBH distribution can attain a range of masses with the following form:
\be
\fBH(M) \propto M^{1.25}\exp\left[\left(\frac{M}{M_c}\right)^{2.85}\right]\,.
\ee
where $M_c$ is the initial peak of the distribution. In the right most plot of \figref{fig:lognormal}, the brown line indicates a typical mass distribution
from such a PBH formation mechanism.\\

\paragraph{Metric Preheating (MP)}~\newline~\newline
Generically, PBHs are expected to form from the collapse of primordial perturbations that may form during or after inflation. In realistic particle physics models, inflation is usually followed by a phase of matter domination where the energy density of the Universe is dominated by the coherent oscillations of the inflaton field. In Ref.~\cite{Martin:2019nuw,Martin:2020fgl,Auclair:2020csm}, it was noted that during that time, perturbations that were generated during the late inflationary era can get resonantly amplified and collapse into black holes before the Universe is reheated. Depending on the reheating temperature, the PBH mass fraction can peak at different masses.  

In order to obtain such a distribution, as shown by the green line in the rightmost plot of  \figref{fig:lognormal}, one should in principle trace the collapse of the oscillating inflaton modes into PBHs numerically. In practice, such a production mechanism leads to an energy fraction of PBHs that is close to one and the PBHs formed dominate the energy density of the Universe quickly. For simplicity, we will consider in what follows the distribution exhibited in the Appendix of Ref.~\cite{Auclair:2020csm}. This distribution (let us denote it by $\fBH^{\rm AV}$) shows a maximum around $M_c^{\rm AV}\sim 10^{5.6}~\mathrm{g}$. Later, to extrapolate these results and explore different mass ranges, we assumed that the overall shape of the distribution does not change for different values of the reheating temperature, and we simply translate the distribution in log-space as follows:
\be
\fBH(M)=\fBH^{\rm AV}\left(M\times\frac{M_c^{\rm AV}}{M_c}\right)\,.
\ee 


\subsection{Spin Distributions}\label{sec:spindistrbutions}

PBHs formed from the collapse of primordial perturbations re-enetering the Hubble horizon are typically created without angular momentum, but may acquire some spin distribution via mergers or during phases of early matter domination \cite{Khlopov:1980mg,Harada:2016mhb,Cotner:2017tir,Cotner:2018vug,Cotner:2019ykd}. 
In particular, if the rate of PBH binary capture is significant compared to the Hubble rate, the in-spiral phase may end before light PBHs start evaporating. After they merge repeatedly, the PBH spin distribution is expected to stabilise to a universal distribution which does not depend anymore on the PBH masses~\cite{Fishbach:2017dwv}.
This spin distribution peaks strongly at $\langle a_\star \rangle\sim 0.7$, with  few BHs with $a_\star< 0.4$. Another spin distribution involving hierarchical mergers was proposed in Ref.~\cite{Doctor:2021qfn} based on LIGO/VIRGO data regarding mergers in the Milky Way. As a matter of fact, this distribution peaks as well at $a_\star\sim 0.7$, and in both cases, the spin distribution is almost entirely independent of the masses or the initial spin distribution of the merging binaries~\cite{Fishbach:2017dwv,Doctor:2021qfn}. 
Although the black holes we consider are much lighter, as compared to the ones used in these different studies, it was shown that two non-rotating BHs with similar masses acquire a spin of order $a_\star\approx 0.69$ after merging~\cite{Hofmann:2016yih}. This universality suggests that the full distribution of PBHs before they start evaporating can be written as a product of the form
\be\label{eq:universality}
\fBH(M,a_\star)=\fBH^M(M)\times\fBH^a(a_\star)\,.
\ee
In what follows, we will assume that this universality property holds for light PBHs produced in the early Universe, and consider the evaporation of distributions that take the form of Eq.~\eqref{eq:universality}. 
Note that this simplifying assumption, and our code is able to handle any general distribution $\fBH(M,a_\star)$.
For concreteness, we will consider either the spin distribution obtained in Ref.~\cite{Fishbach:2017dwv} or simply take a Gaussian distribution centered around $a_\star = 0.7$ and having width $\sigma_{a_\star}$. Other mechanisms of PBH production, such as Q-balls, oscillons, or cosmic strings, may as well be created with specific spin distributions~\cite{Cotner:2016cvr,Cotner:2017tir,Cotner:2019ykd,Cotner:2016cvr,Cotner:2017tir,Cotner:2019ykd,Cotner:2018vug,Jenkins:2020ctp}. However, giving a comprehensive review of the effect of different distributions is beyond the scope of this work. In this paper, we focus on the spin distribution from hierarchical mergers as a simple case study, to exhibit the general phenomenological effects of spin distributions. 

\section{Primordial Black Hole Evaporation}\label{sec:evaporation}

In this section, we review the physics of PBH evaporation. For the sake of generality, we consider the case of a Kerr PBH, that is characterized by both its mass and angular momentum. The evaporation of a Schwarzschild PBH will simply correspond to the special case of vanishing angular momentum. 
In what follows, we do not consider effects from dynamical horizons on the particle production given the lack of consensus on which metric is the correct one to describe cosmological black holes, see~\cite{Barrau:2022bfg} and references therein for the status of this topic.
In other words, we assume that the Hawking evaporation operates in the same way as in vacuum.

We denote  the mass of a PBH and its dimensionless spin parameter by $\MBH$  and $a_\star=J\MPL^2/\MBH^2$ respectively, where $J$ is the angular momentum of the black hole. The Hawking emission rate for a particle species $i$ with three-momentum $p$, total energy $E_i$ and $g_i$ internal degrees of freedom  is given by~\cite{Hawking:1974rv,Hawking:1974sw}
\begin{align}\label{eq:KBHrate}
\frac{\dd^2 \mathcal{N}_{i}}{\dd p\dd t}&=\frac{g_i}{2\pi^2} \sum_{l=s_i}\sum_{m=-l}^l\frac{\dd^2 \mathcal{N}_{ilm}}{\dd p\,\dd t}\,,
\end{align}
where
\begin{align}\label{eq:KBHrate_lm}
\frac{\dd^2 \mathcal{N}_{ilm}}{\dd p\dd t}&=\frac{\sigma_{s_i}^{lm}(\MBH,p,a_\star)}{\exp\left[(E_i - m\Omega)/\TBH\right]-( -1)^{2s_i}}\frac{p^3}{E_i}\,.
\end{align}
In this expression, $\Omega = (a_\star/2G\MBH)(1/(1+\sqrt{1-a_\star^2}))$ is the horizon's angular velocity, and $l,m$ are the total and axial angular momentum quantum numbers, respectively.
In Eq.~\eqref{eq:KBHrate}, it is necessary to  write explicitly the sum over the angular momentum quantum numbers as the PBH spin breaks the spherical symmetry, given the explicit dependence of the Hawking rate on $m$. 
For the case of Schwarzschild PBHs, $a_\star = 0$, the emission rate is independent of $m$ so one can replace the sum over $m$ by a factor of $\left(2l+1\right)$.
$\sigma_{s_i}^{lm}$ appearing in Eq.~\eqref{eq:KBHrate} corresponds to the BH absorption cross section, which describes the effects of the centrifugal and gravitational potential on the particle emission~\cite{Hawking:1974rv,Hawking:1974sw,Page:1976df,Page:1976ki,Page:1977um}.
These quantities are determined by solving the spin-dependent equations of motion for a field in the Kerr spacetime.
We have adopted the methodology described in Refs.~\cite{Chandrasekhar:1975zz,Chandrasekhar:1976zz,Chandrasekhar:1977kf} to obtain such cross-sections.

To derive evolution equations for the BH mass and spin, we multiply the Hawking rate shown in Eq.~\eqref{eq:KBHrate} by the total energy of a given particle $E_i$ or by the $m$ quantum number, and then we integrate over the phase space. 
Defining the evaporation functions for mass and angular momentum, $\varepsilon_i(\MBH, a_\star)$ and $\gamma_i(\MBH, a_\star)$ per particle $i$, respectively, as
\begin{align}
 \varepsilon_i(\MBH, a_\star) &= \frac{g_i}{2\pi^2}\int_{0}^\infty \sum_{l=s_i}^\infty\sum_{m=-l}^l\frac{\dd^2 \mathcal{N}_{ilm}}{\dd p\dd t}\,E dE\,,\\
 \gamma_i(\MBH, a_\star) &= \frac{g_i}{2\pi^2}\int_{0}^\infty \sum_{l=s_i}^\infty\sum_{m=-l}^l m \frac{\dd^2 \mathcal{N}_{ilm}}{\dd p\dd t}\, dE\,,
\end{align}
where $g_i$ is the particle species $i$'s internal degree of freedom and we can  sum over \emph{all} existing species to obtain the following system of coupled equations~\cite{PhysRevD.41.3052,PhysRevD.44.376,Cheek:2021odj}
\begin{subequations}\label{eq:dynamicsKerr}
\begin{align}\label{eq:dynamicsevaporation}
 \frac{d \MBH}{dt} &= - \varepsilon(\MBH, a_\star)\frac{M_p^4}{\MBH^2}\,,\\
 \label{eq:dynamicsspin}\frac{da_\star}{dt} &= - a_\star[\gamma(\MBH, a_\star) - 2\varepsilon(\MBH, a_\star)]\frac{M_p^4}{\MBH^3}\,.
\end{align}
\end{subequations}
We refer the interested reader for further details of the derivation of these equations to Refs.~\cite{Cheek:2021odj, Cheek:2022dbx}. In this equation, we introduced the total evaporation functions for mass and angular distribution,
\be
\varepsilon \equiv \sum_i \varepsilon_i\,,\quad\text{and}\quad\  \gamma\equiv\sum_i\gamma_i\,.
\ee
Later in the paper, we will use similar notations with additional subscripts `SM', `DM', and `DR', indicating that the sum over $i$ is restricted  respectively to Standard Model, dark-matter, and dark-radiation species only.

In general, obtaining the time evolution of a Kerr PBH mass and spin parameters  requires using  numerical methods. 
In our code, we solve numerically the system of equations Eqs.~\eqref{eq:dynamicsKerr} including the dependence of the evaporation functions on the mass of the emitted particles. 
Thus, our approach can be readily extended to include any number of degrees-of-freedom. For instance, the production of an extended sector having a large number of degrees-of-freedom, as explored in e.g. Refs.~\cite{Calza:2021czr,Baker:2021btk}, could be included which could lead to interesting phenomenology. 

For the sake of numerical efficiency, we will, however, restrict our analysis to evaporation products with masses smaller than the initial PBH Hawking temperatures. Note that in the case where the evaporation functions $\varepsilon$ and $\gamma$ only depend on the PBH spin, that is, in the limit where the evaporation products can be considered to be exactly massless, there is a formal solution of these equations, as first described in Ref.~\cite{Page:1976ki}. For completeness, we briefly revisit this derivation in the App.~\ref{ap:KPBH_sol}.

\section{Evaporation of an Extended Distribution}\label{sec:Schwarz_BHs}
In this section, we derive the Boltzmann and Friedmann equations that allow tracking the evolution of PBH abundance and the Universe's energy density throughout cosmic history, for an evaporating distribution of PBH. We start by defining $\fBH(\MBH,a_\star, t)$ as the distribution of PBH of mass $\MBH$, spin $a_\star$ at time $t$. The number density of PBHs, $n_{\rm BH}$, is then given by
\be
n_{\rm BH}(t)=\int_0^1\int_{0}^\infty \fBH(\MBH, a_\star, t) d \MBH d a_\star\,.
\label{eq:def_nBH}
\ee
Similarly, one can write the comoving energy density of the Black Hole distribution as
\be
\rho_{\rm BH}(t)=\int_0^1\int_0^\infty \MBH \fBH(\MBH, a_\star, t) d \MBH d a_\star\,.
\label{eq:def_rhoBH}
\ee
Demanding that the comoving number of PBHs is conserved throughout cosmic history\footnote{{Note that, when referring to the total comoving number density of PBHs, we include the number of PBHs that have already evaporated, and thus have mass zero. This allows us to track the evolution of the PBH population as a whole without qualitative distinction.}}, we obtain the continuity equation 
\be
3H\fBH = -\frac{\partial{\fBH}}{\partial{\MBH}}\frac{d \MBH}{d t} -\frac{\partial{\fBH}}{\partial{a_\star}}\frac{d a_\star}{d t} - \frac{\partial{\fBH}}{\partial{t}},
\label{eq:dist_relation}
\ee
$H$ being the Hubble parameter.
Taking the time derivative of Eq.~\eqref{eq:def_rhoBH} and using the relation in \equaref{eq:dist_relation} (see App.~\ref{ap:KPBH_sol}) one obtains the Friedmann-Boltzmann equation
\be\label{eq:BoltzmannPBH}
\dot{\rho}_{\rm BH}+3H\rho_{\rm BH} = \int_0^1\int_0^\infty \frac{d\MBH}{d t}\fBH\,d \MBH d a_\star\,,
\ee 
which has to be solved simultaneously with the equation describing the evolution of the Standard Model plasma, $\rho_{\rm SM}$,
\be\label{eq:BoltzmannSM}
\dot{\rho}_{\rm SM} + 4 H \rho_{\rm SM} = - \int_0^1\int_0^\infty\frac{\varepsilon_{\rm SM}}{\varepsilon}\frac{d\MBH}{dt} \fBH\,d\MBH d a_\star\,.
\ee
To solve these equations numerically, one needs to evaluate, at every time $t$, the integrals in the right-hand sides of Eqs.~\eqref{eq:BoltzmannPBH} and \eqref{eq:BoltzmannSM}. However, the distribution $\fBH(M)$ is only known at the time of PBH formation $t_{\rm in}$. In order to obtain the values of these integrals, it is useful to change variables and map the mass spectrum at time $t$ to the corresponding initial masses $\MBHi$, defined such that
\be
(t_{\rm in}, \MBHi, \asi)\longrightarrow(t,\MBH, a_\star)\,.
\ee 
In that case, the conservation of the infinitesimal PBH comoving number density\footnote{The number of PBHs with mass and spin within the range $[M,M+dM]$ and $[\as,\as+d\as]$.} provides that
\bea
\label{eq:conditionM}
a^3(t)d n_{\rm BH} &\equiv& a^3(t)\fBH(\MBH,\as,t)d \MBH d \as\nonumber\\
&=& a^3(t_{\rm in})\fBH(\MBHi, \asi,t_{\rm in})d \MBHi d \asi\,,\nonumber\\
&\equiv & \mathcal F_{\rm in}d \MBHi d \asi
\eea
 where $a$ stands for the scale factor. For future convenience we defined $\mathcal F_{\rm in}\equiv a^3(t_{\rm in})\fBH(\MBHi, \asi,t_{\rm in})$. The Boltzmann equations  that are solved, after defining $\varrho_{\rm BH}\equiv a^3\rho_{\rm BH}$ and 
$\varrho_{\rm SM}\equiv a^4\rho_{\rm SM}$, are

\bea\label{eq:BoltzmannPBHandSM}
\dot{\varrho}_{\rm BH} &=& \int\int \frac{d\MBH}{d t}\mathcal F_{\rm in}\,d \MBHi d\asi\,,\nonumber\\
\dot{\varrho}_{\rm SM}&=& - a(t) \int\int\frac{\varepsilon_{\rm SM}}{\varepsilon}\frac{d\MBH}{dt} \mathcal F_{\rm in}\,d\MBHi d\asi\,.
\eea
Note that in these expressions, $d\MBH/dt$ remains a function of $\MBH=\MBH(t,\MBHi, \asi,t_{\rm in})$ and $\as=\as(t,\MBHi, \asi,t_{\rm in})$ which can be found numerically by integrating Eqs.~\eqref{eq:dynamicsKerr}. Moreover, one should pay attention to the fact that at the end of the evaporation $d\MBH/dt$ diverges, and thus the mass integrals in the right-hand sides of Eqs.~\eqref{eq:dynamicsKerr} have to be restricted to initial masses that have not yet reached 0 at time $t$.

\section{Numerical Implementation}
\label{sec:numerical}
Numerically, to solve the system of Eq.~\eqref{eq:BoltzmannPBHandSM} we adopt the following procedure:
\begin{enumerate}
    \item For a given particle physics model, i.e., specifying the particle content, calculate the evolution of a close-to-maximally rotating PBH by numerically solving the system in Eqs.~\eqref{eq:dynamicsKerr}, which will be required for the determination of the evolution of any other PBH~\cite{Cheek:2022dbx}.
    \item Determine the PBH lifetime for a fixed mass $M_{\rm fix}$ as function of the spin parameter $\asi$. With this, we can determine the lifetime of any PBH using the relation
    \begin{align}\label{eq:relativisticdof}
        \tau(\MBHi,\asi) \approx \tau(M_{\rm fix}, \asi) (\MBHi/M_{\rm fix})^3.
    \end{align}
    \item Fixing the time $t$ and a set $(\MBHi,\asi)$, calculate the value of $M(t)$ and $a(t)$.
    \item Evaluate numerically the corresponding evaporation rate $dM/dt(M(t),a(t))$, as well as the greybody factors $\varepsilon(M(t))$ and $\varepsilon_{\rm SM}(M(t))$.
    \item Integrate over $(\MBHi,\asi)$ restricting the 2D integration volume of Eq.~\eqref{eq:BoltzmannPBHandSM} to initial masses and spins that satisfy $\tau(\MBHi,\asi)\geqslant t$. { Indeed, we assume that BHs with shorter lifetimes have fully evaporated at time $t$ and do not contribute neither to the comoving energy density of PBHs, nor to its variation with time written in Eq.~\eqref{eq:BoltzmannPBHandSM}\footnote{{Note that if PBHs were assumed to stop evaporating when their mass equals the Planck mass, then one would need to keep track of Planck relics when computing the comoving energy density. Similarly, the evaporation rate $dM/dt$ should be set to zero when $M=M_p$.}}}.
\end{enumerate}

\begin{figure*}
\centering
    \includegraphics[width=\linewidth]{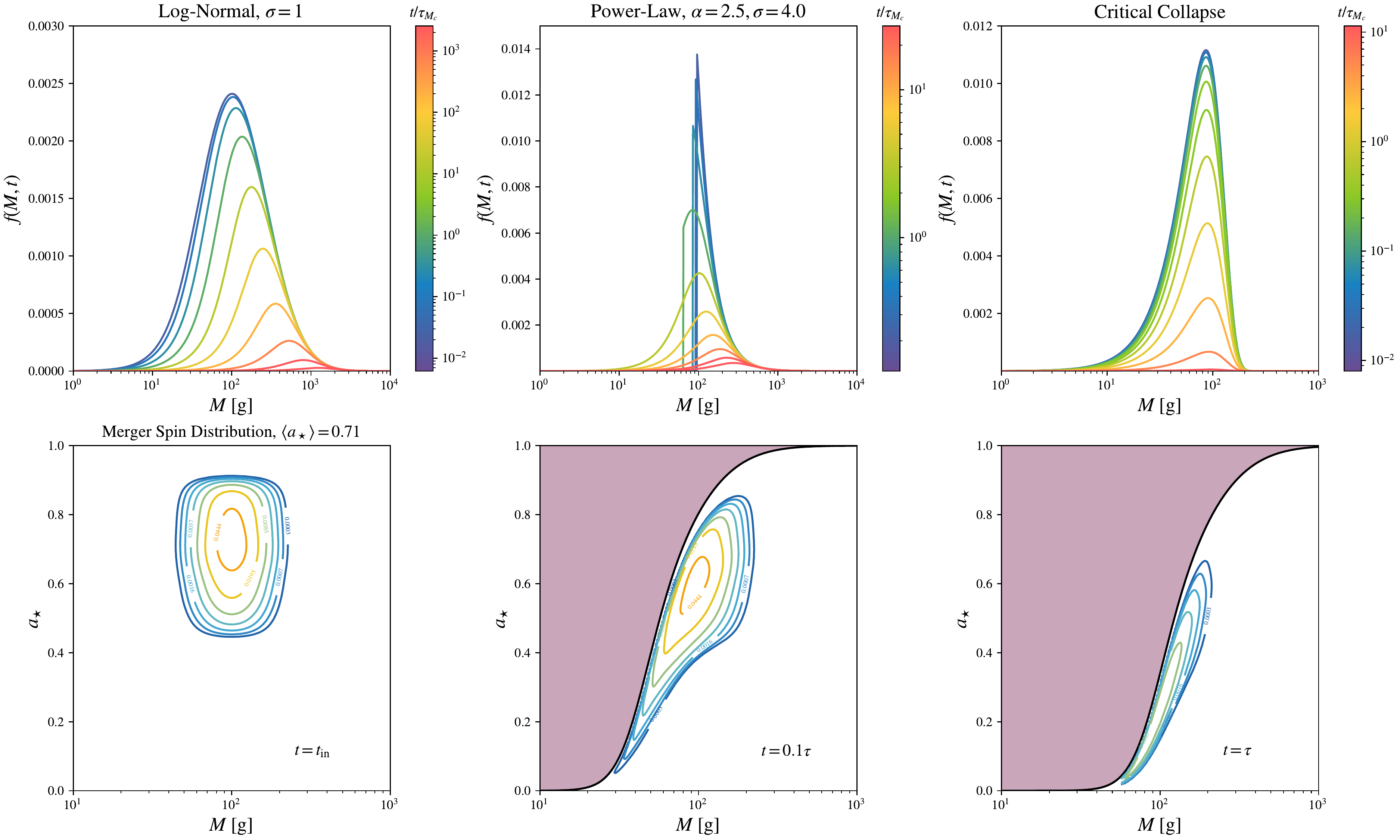}
    \caption{\label{fig:time_evol}\footnotesize 
    Top panels: Time evolution of the comoving mass distribution $f\equiv a^3(t)f_{\rm PBH}$ for a log-normal (left panel), power-law (central panel), and critical collapse (right panel) distributions. The log-normal distribution is taken to have a width $\sigma=1$ and the central mass considered is $M_c=10^{2}\mathrm{g}$. $\tau_{M_c}$ indicates the lifetime of a PBH with a mass equal to $M_c$.
    Bottom panels: Evolution of a mass and spin distribution $\fBH(\MBH, \as, t)$ as a function of time, starting with a log-normal distribution in mass, and the universal spin distribution obtained in Ref.~\cite{Fishbach:2017dwv}. The left (right) most plot shows the initial (final) distribution at time an initial (final) time. The coloured lines are iso-contours of the distribution, whereas the lifetime $\tau$ denotes the lifetime of the PBH with initial mass and spin at the peak of the distribution. The shaded region represents the region where PBHs would have a lifetime shorter than the corresponding time $t-t_{\rm in}$, and can therefore not be present in the Universe at time $t$.}
\end{figure*}
We assume that the PBH distribution is formed simulatenously when the Universe is radiation dominated after inflation, and we take as formation time $t_{\rm in}$ --or temperature $T_{\rm in}$-- when the particle horizon mass is equal to mass of the initial peak of the distribution $M_c$
\begin{align}
    M_c = \frac{4\pi}{3}\kappa \frac{\rho_R(T_{\rm in})}{H^3(T_{\rm in})},
\end{align}
with $\kappa\sim 0.2$. 
We stress that this is a simplifying assumption since it is expected that each formation mechanism will predict different initial conditions, such as $t_{\rm in}$.
We note however that our code can be easily modified to specify the initial conditions and the initial time $t_{\rm in}$. Alternatively, if one had a PBH production mechanism that continued to form black holes while some smaller PBHs were evaporating, more substantial modifications of {\tt FRISBHEE} would be required.   
The double integration is performed at every time step in the range $t\in [t_{\rm in}, t_{\rm fn}]$, $t_{\rm fn}$ being the time when the whole PBH population has evaporated. We then track the evolution of the PBH population and the expansion dynamics of the Universe by solving the Friedmann and Boltzmann equations of Eq.\eqref{eq:BoltzmannPBHandSM}, making use of the pre-calculated 2D integrals.

We have implemented this numerical strategy in the publicly available code\footnote{Available at \href{https://github.com/yfperezg/frisbhee}{https://github.com/yfperezg/frisbhee}. We provide codes for scenarios with mass only and mass and spin distributions. The mass only code follows the same established procedure detailed above, although it does not uses the approximated PBH lifetime instead using its full numerical value.} {\tt FRISBHEE} \href{https://github.com/yfperezg/frisbhee}{\faGithub} which follows these steps and is able to include SM and BSM particles in the Hawking evaporation spectrum considered. We have double-checked that we recover the results from Refs.~\cite{Cheek:2021cfe,Cheek:2022dbx,Cheek:2021odj} when we take the monochromatic limit for a given mass and/or spin distribution. To our knowledge, this is the first implementation of mass and spin distributions into the Boltzmann and Friedmann of the early Universe. Although the tool, {\tt BlackHawk}~\cite{Arbey:2019mbc, Arbey:2021mbl} includes the facility to compute Hawking radiation from a distribution of black holes, it does not solve the cosmic evolution alongside BH evaportation. That's because {\tt BlackHawk}'s main focus is the accurate determination of primary and secondary spectra from black holes that are currently evaporating, $\MBH\sim10^{14}\,{\rm g}$, which with existing constraints will simply be a background matter density in the early Universe.

Note that the relation we used in Eq.~\eqref{eq:relativisticdof} does not hold in full generality. It is true to very good accuracy for evaporation products that are much lighter than the Hawking temperatures of the PBHs in the distribution. However, a PBH's lifetime starts deviating from its true value as soon as a large number of new degrees of freedom with masses larger than the Hawking temperature of the PBH are introduced in the spectrum. As a matter of fact, when adding 200 new degrees of freedom with masses $m\ll \TBH$, this equation is verified to 0.005\% accuracy. It reduces to a 30\% error for $m>\TBH$. For the sake of numerical efficiency, we have therefore implemented this approximation in {\tt FRISBHEE} to improve its calculation speed. 
However, we stress that employing the lifetime of PBHs for any interval $[M,M+dM]\times [a_\star,a_\star+da_\star]$, which our code is capable to perform, would not require a significant modification, although it would imply a larger run time.

It is possible to trace the evolution of the distribution at any time by computing the relevant Jacobian,
\be
a^3(t)\fBH(\MBH, \as, t) = a^3(t_{\rm in})\frac{\fBH(\MBHi,\asi, t_{\rm in})}{\mathcal J}\,.
\ee
As described in Appendix \ref{ap:KPBH_sol}-\ref{ap:Nconservation}, in the case of relativistic evaporation products, we find an explicit expression for the Jacobian
\be
\mathcal J \equiv  \left|\frac{\partial \MBH}{\partial \MBHi}\frac{\partial \as}{\partial \asi}-\frac{\partial \MBH}{\partial \asi}\frac{\partial \as}{\partial \MBHi}\right|\,.
\ee
We present in Fig.~\ref{fig:time_evol} an example of the time evolution of the mass only and mass and spin distributions as function of time.
In the case of only mass distribution (top panels), we present the time evolution of a log-normal (left), power-law (middle) and critical collapse (right) scenarios, for different values of time, as indicated in the figures by the different color scale. Note that we recover, for the power-law distribution, the results derived analytically in Ref.~\cite{Barrow:1991dn}.
In the case of a mass and spin distribution (lower panels) we present three different snapshots for the initial time $t_{\rm in}$ (left), a time equal to 10\% of the lifetime of the peak $\tau$ (middle) and $t=\tau$ (left). 
The shaded region represents the PBH parameters which cannot be present in the Universe since such PBHs would have a lifetime shorter than $t-t_{\rm in}$. Note that, although it looks like all these distributions get depleted with time, that the comoving number density of PBHs (including those whose masses is zero after evaporation) is conserved in such simulations. Indeed, the evaporation of a black hole is a process that accelerates with time and, therefore, the mass of PBHs that start evaporating quickly runs towards zero. Numerically, the depletion of the distributions that is visible in the figure is balanced by a delta function for mass and spin located at the origin and increasing with time. Note that in practice, {\tt FRISBHEE} stops tracking the evolution of the PBH mass once it falls below 10\% of the Planck Mass and sets it to zero, but keeping and tracking Planck relics in the code is possible.

We have included some animations of the time evolution of mass only and mass and spin distributions for benchmark parameters. These are included in the wiki\footnote{ \href{https://github.com/yfperezg/frisbhee/wiki/Animations-Mass-and-Spin-Distributions}{https://github.com/yfperezg/frisbhee/wiki/Animations-Mass-and-Spin-Distributions}} page of {\tt FRISBHEE}'s GitHub.

\section{Cosmological Imprints of PBH Distributions}
\label{sec:observables}

In this section, we explore the possible consequences that the evaporation of an extended distribution of PBHs can have in cosmology, and how one can attempt to probe the shape of such a distribution using cosmological observables. 
In Ref.~\cite{Cheek:2021cfe, Cheek:2021odj, Cheek:2022dbx}, it was shown that the evaporation of PBHs can leave various imprints in dark-matter searches, small-scale structures, and dark radiation measurements. In the next sections, we will see how the modification of the cosmological background's evolution arising from the extension of the PBH distribution can affect such observables.

In fact, it is expected that any mechanism that involves the production of particles out of equilibrium (such as e.g. leptogenesis, see Ref.~\cite{Perez-Gonzalez:2020vnz}) throughout cosmic history can be affected by such dynamics. Gravitational waves induced at second order in perturbation theory constitute interesting signatures of a possible PBH dominated era. Arising from the successive onset of the PBH domination and PBH evaporation, they could also reveal interesting signatures regarding the shape of the PBH distribution~\cite{Bhaumik:2022pil}. We will briefly discuss this possibility, and leave an extensive study of these different imprints for future works.
\begin{figure*}[t!]
    \centering
    \includegraphics[width=0.33\linewidth]{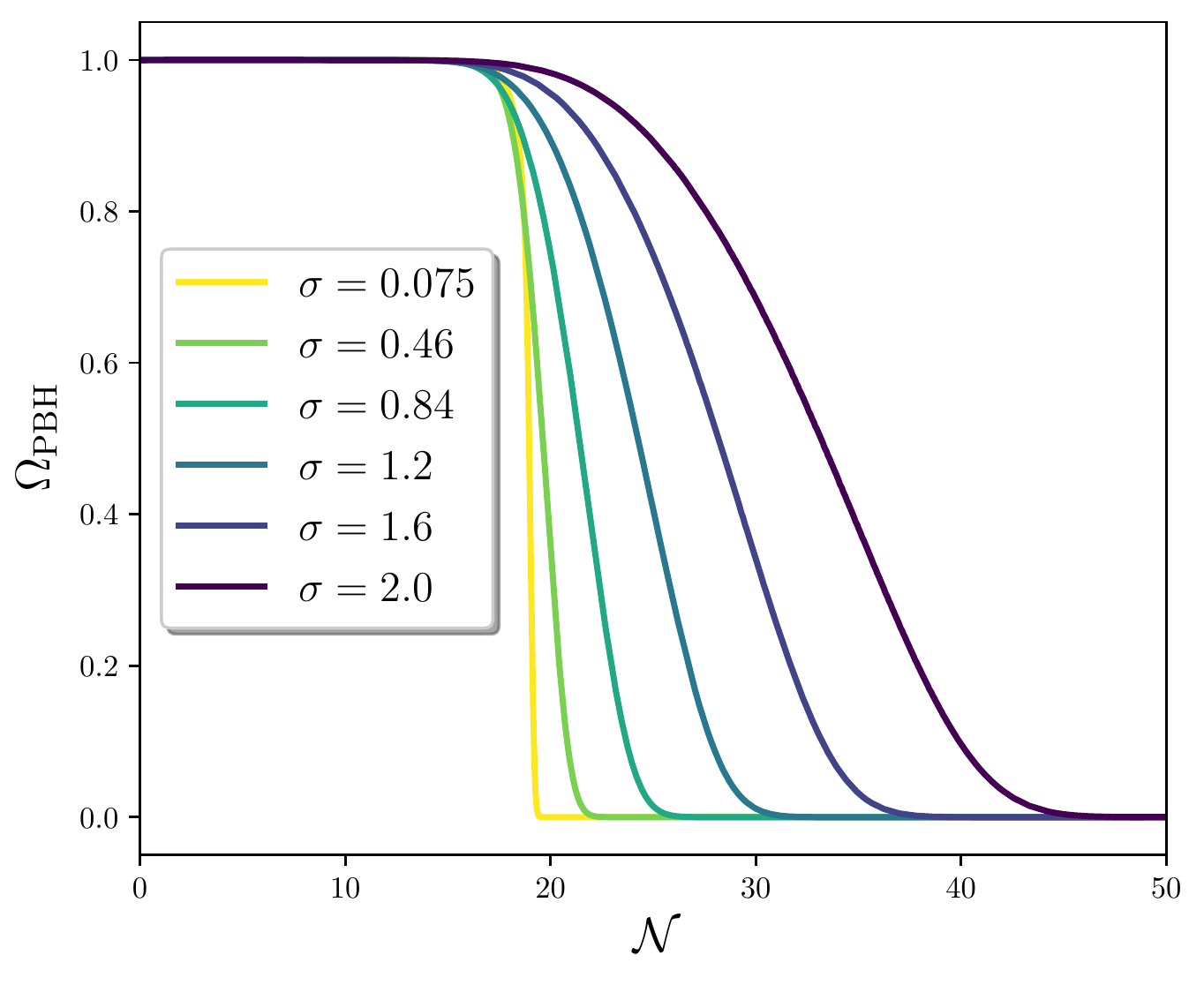}\includegraphics[width=0.33\linewidth]{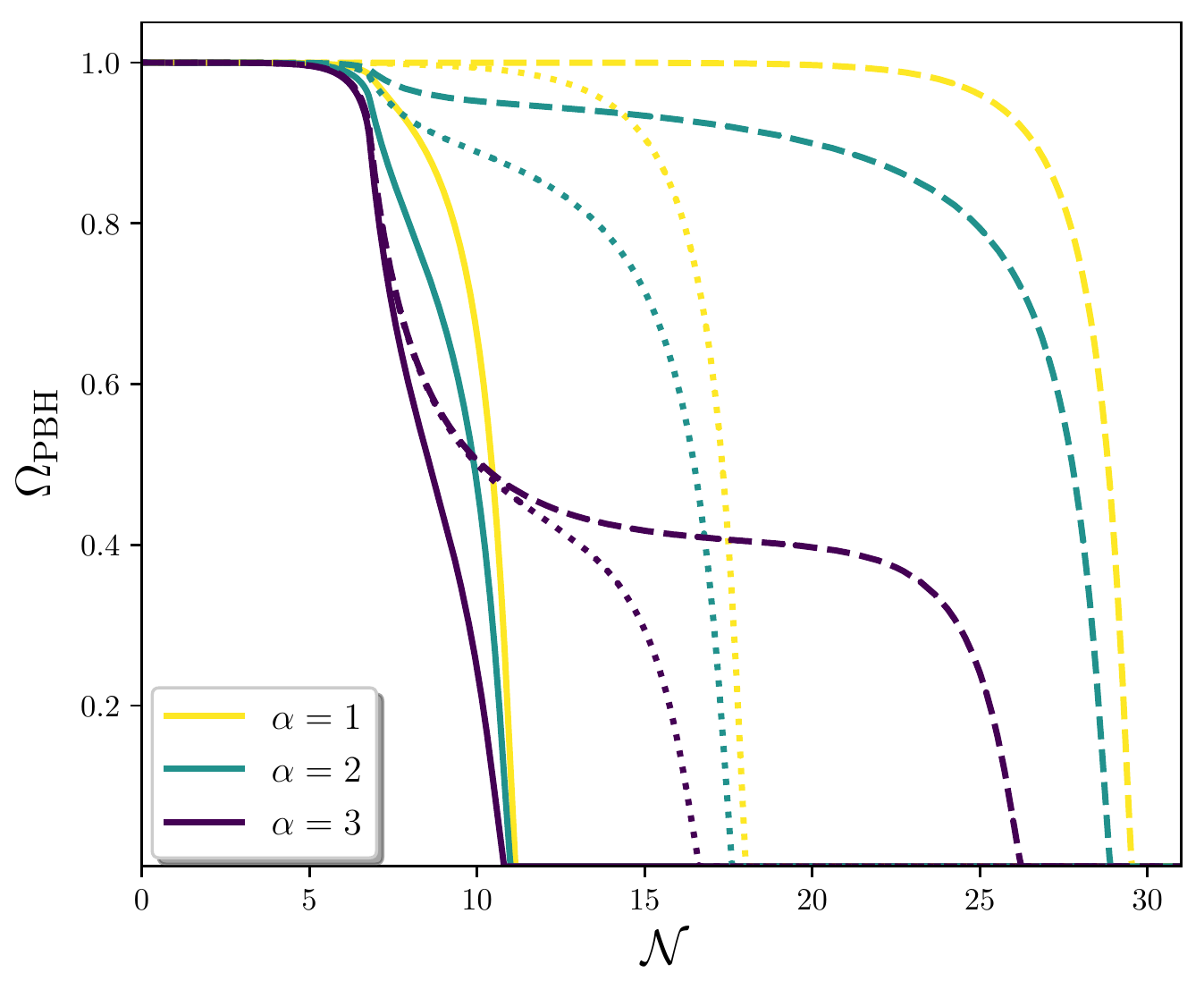}\includegraphics[width=0.33\linewidth]{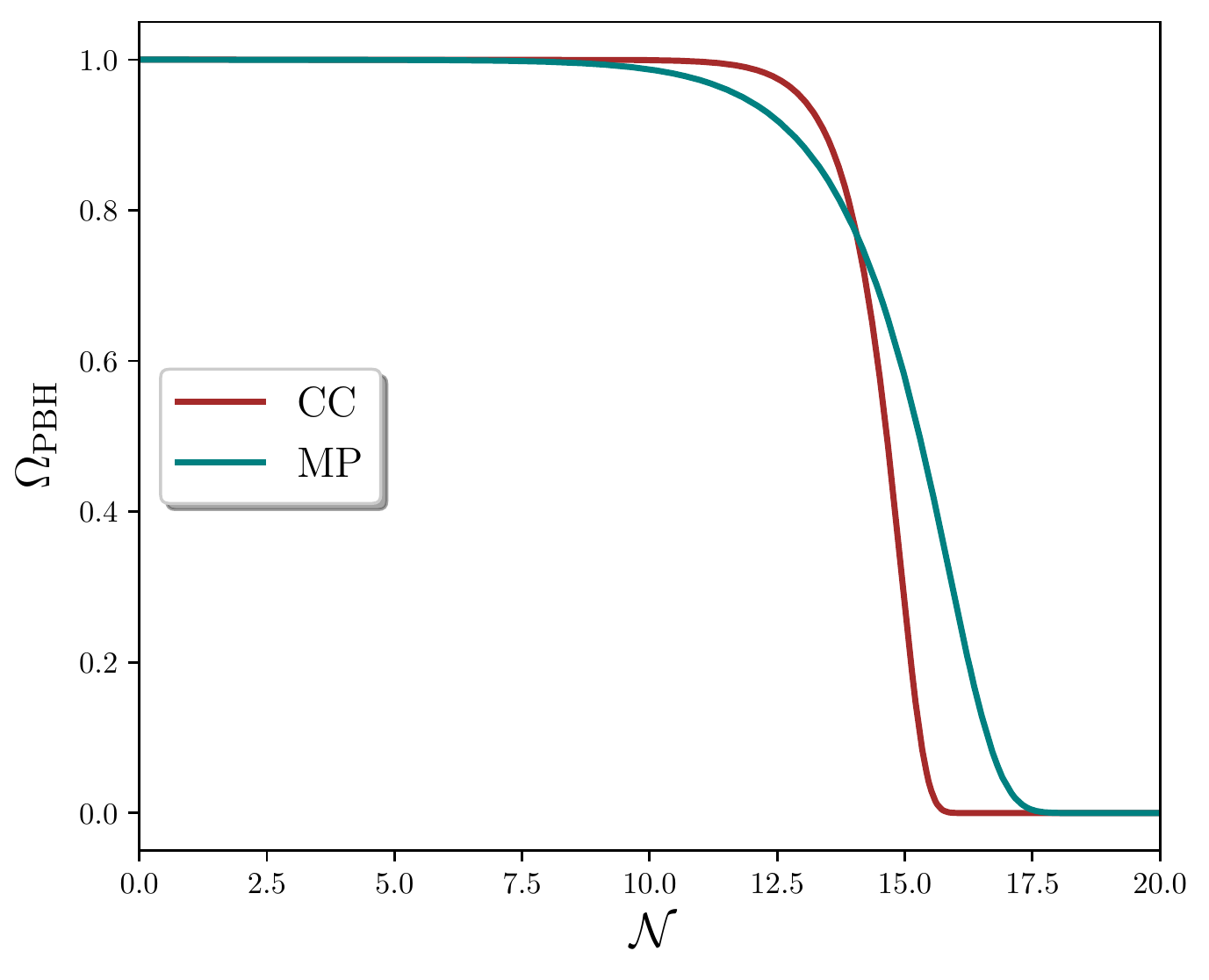}
    \caption{\label{fig:CriticalAndVennin}  \footnotesize The right plot shows the evolution of the PBH relative abundance $\Omega_{\rm PBH}$ as a function of the number of  $e$-folds $\mathcal N$ for log-normal distributions of PBHs with varying widths $\sigma$. The central plot shows the same but for power-law distributions of PBHs with varying exponents $\alpha$, and minimum mass $M_c=10~\mathrm{g}$. Plain, dotted, and dashed lines stand for the evaporation of distributions with, respectively $M_c 10^\sigma= 10,\ 10^{2.5},\ $and $10^5\mathrm{g}$. The right plot shows the PBH evolution as a function of the number of $e$-folds $\mathcal N$ for the distributions from critical collapse and metric preheating (see Sec.~\ref{sec:distributions}).}
\end{figure*}

\subsection{PBH Domination}\label{sec:cosmo}

In the case of a monochromatic distribution, $\fBH(\MBH)=\delta(\MBH-\MBHi)$, PBH domination can be assimilated to an early matter-domination period that ends when the whole population of PBHs evaporates.
Such sudden evaporation can have many consequences in cosmology, and possibly leave observable imprints in cosmological data. The existence of a PBH dominated era typically affects the dynamics of the Universe expansion as compared to the temperature evolution of the SM sector, as well as the time-evolution of the Universe's equation of state parameter. 
If primordial black holes form in the early Universe with a sufficient energy fraction, they can end up dominating the energy density of the Universe before they evaporate. When this is the case, their evaporation corresponds to an injection of entropy into the SM sector that can strongly affect its dynamic with respect to the Universe's expansion.  In this section, we explore the effect that the extended PBH distributions have on the dynamics of evaporation and corresponding entropy injection into the SM bath. For the sake of simplicity, this section only considers Schwarzschild PBHs. Note that the presence of a spin distribution would however affect the following results, as spinning PBHs would have the tendency of evaporating faster. However, we believe that the qualitative dynamics of reheating that
we present are not to be changed drastically. Indeed, the universality of the spin distribution considered in Eq.~\eqref{eq:universality} guarantees that the lifetime of PBHs with different masses would be shortened in a relatively uniform way. Besides happening at a slightly earlier time, the qualitative dynamics of the Universe's reheating when PBHs evaporate would not be significantly affected by the shape of the spin distribution.

In Fig.~\ref{fig:CriticalAndVennin}, we present our results for the different mass distributions listed in \secref{sec:distributions}. We draw the evolution of the PBH relative abundance $\Omega_{\rm PBH}\equiv \rho_{\rm PBH}/\rho_{\rm tot}$ as a function of the number of $e$-folds $\mathcal N\equiv \log a$, with $a$ denoting the scale factor. In the case of a log-normal distribution (left panel), we vary the width $\sigma$, introduced in Eq.~\eqref{eq:distri_lognormal}, and we fix the central mass $M_c$ to be $10^4$ g.  In the small-$\sigma$ limit, one can observe that evaporation happens quickly, similar to the case of a monochromatic distribution. On the other hand, increasing $\sigma$ corresponds to smearing the evaporation over larger time scales, and one can see that the evaporation can then happen over several $e$-folds of expansion. A similar observation can be made in the case of the power-law distribution introduced in Eq.~\eqref{eq:distri_powerlaw}. In the central plot of Fig.~\ref{fig:CriticalAndVennin}, we used different powers $\alpha$ corresponding to the formation of PBHs during matter domination ($\alpha=2$), kination ($\alpha = 0$), and a nearly inflating Universe ($\alpha\gtrsim 1$). We fixed the mass $M_c=10$ g and varied the width of the distribution by taking $M_c\times 10^{\sigma}=10^2{\rm g},10^{3.5}{\rm g}, \text{ and }10^6{\rm g}$. As expected, increasing the width of the distribution leads to extending the duration of the evaporation process.

It is interesting to note that for certain choices of $\alpha$ (namely $1<\alpha\leqslant 3$), as shown in the middle panel of Fig.~\ref{fig:CriticalAndVennin},
the total black hole abundance initially falls but then remains essentially fixed at a non-zero value over an interval lasting many $e$-folds before finally dropping to zero. It turns out that this feature is a specific instance of a more general phenomenon called {\em cosmic stasis}\/~\cite{Dienes:2021woi}, wherein the abundances of the different energy components of the Universe remain fixed over an extended interval despite cosmological expansion.
In this case, the total PBH abundance is remaining fixed because its natural tendency to grow in this Universe as a result of redshifting effects is precisely counterbalanced by the loss of PBH energy density into radiation via Hawking evaporation~\cite{Barrow:1991dn, Dienes:2022zgd}. It is remarkable that distributions with power $1<\alpha\leqslant 3$ are all attracted to such a regime.

Finally, in the rightmost plot of  Fig.~\ref{fig:CriticalAndVennin}, we illustrate the evaporation dynamics which is obtained when using the distributions corresponding to the critical collapse and metric preheating cases, as described in Sec.~\ref{sec:distributions}. Here both distributions peak towards heavy masses, and thus the evaporation process is dominated by the end tail of the distribution.


\subsection{Dark Matter Relic Density}
\label{sec:DM}
In this section, we investigate the impact of the evaporation of a PBH population with both mass and spin extended distributions on DM production.
Besides the equations for the evolution of the SM radiation and PBH energy densities, we need to track the number density of DM particles produced from the evaporation.
Here, we assume that such particles only interact gravitationally, and, although there might be other production mechanisms, we focus only on the DM from Hawking evaporation.
Similarly to the Eqs.~\eqref{eq:BoltzmannPBH}, we can write the equation for the DM comoving number density $N_{\rm DM}$ as 
\begin{align}
    \dot{N}_{\rm DM} = - \int\int\frac{\varepsilon_{\rm DM}}{\varepsilon}\frac{d\MBH}{dt} {\cal F}_{\rm in}\,d\MBHi d\asi\,.
\end{align}
where now the evaporation function includes the contribution of the DM particle, $\varepsilon_{\rm DM}$. 
At this point, we do not fix the spin of the particle, to make our discussion as general as possible.
In order to compare with previous results that consider a monochromatic distribution, we  parametrize the initial PBH energy density via 
\begin{align}
 \beta^\prime \equiv \kappa^{1/2}\left(\frac{g_\star (T_{\rm in})}{106.75}\right)^{-1/4}\frac{\rho^{\rm in}_{\rm PBH}}{\rho_{\rm rad}^{\rm in}}\,,
\end{align}
where $\rho_{\rm PBH}^{\rm in}$ is related to the mass and spin distribution as in Eq.~\eqref{eq:def_rhoBH}.
\begin{figure*}[tb]
    \centering
    \includegraphics[width=.45\linewidth]{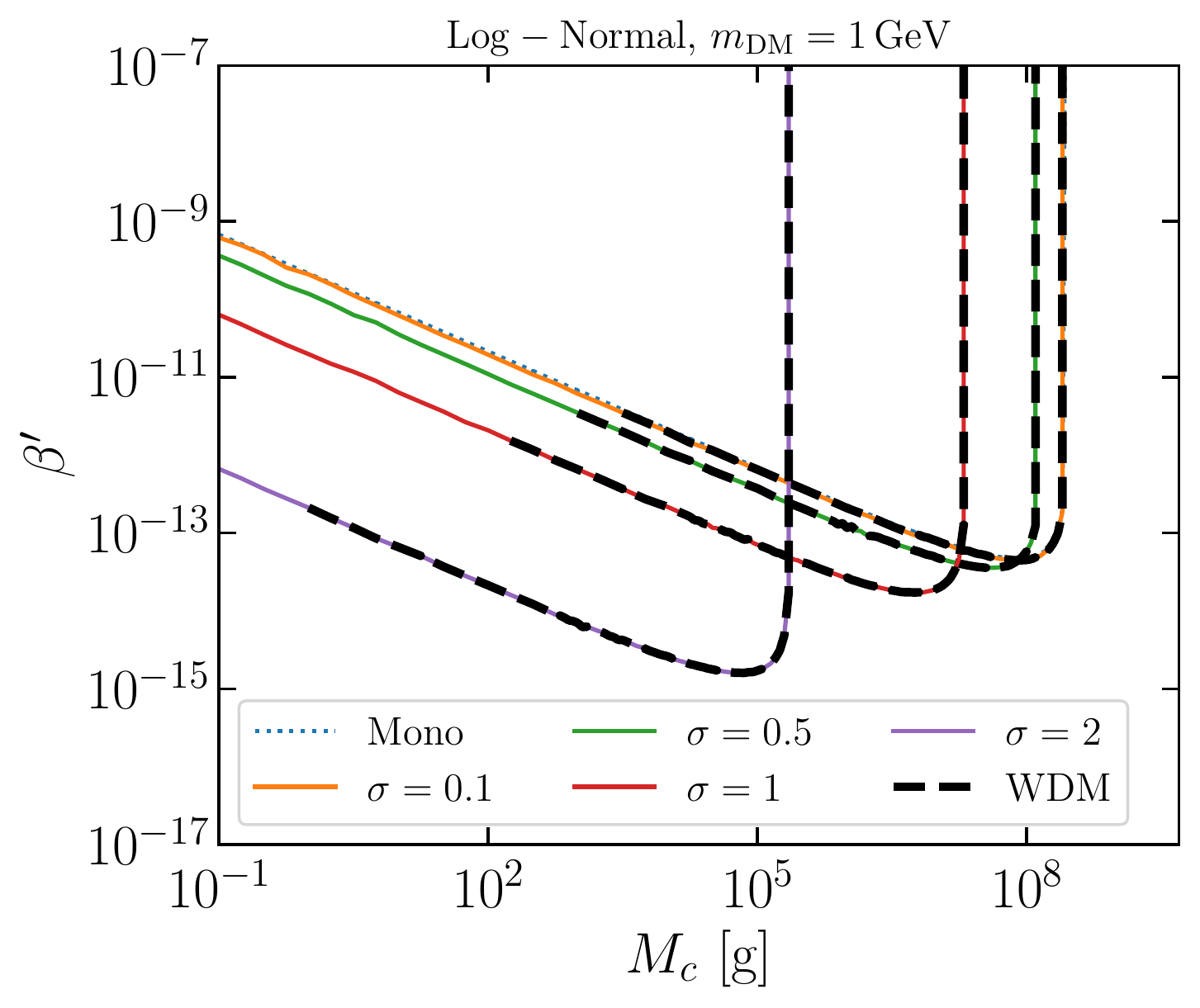}
    \includegraphics[width=.45\linewidth]{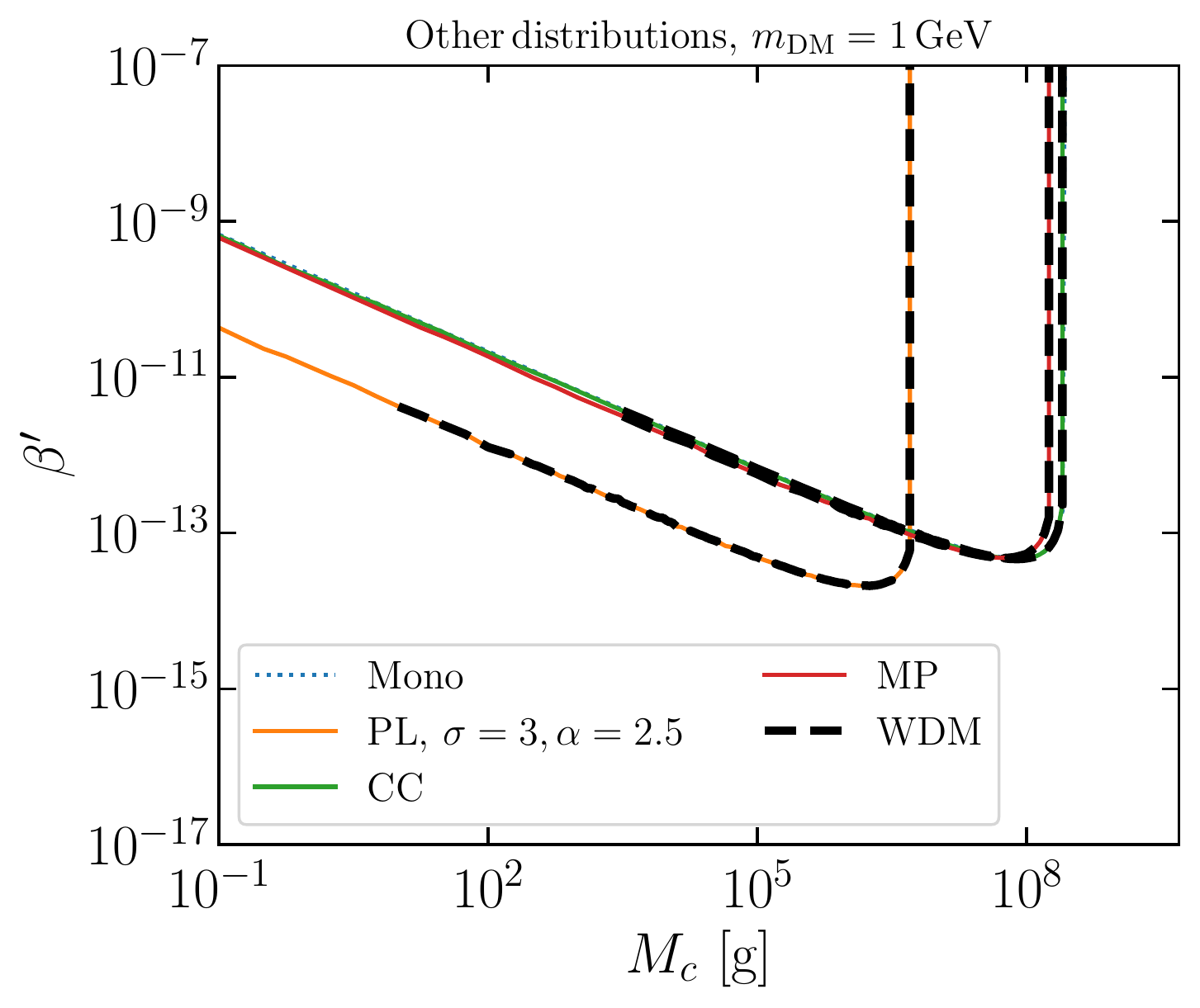}
    \caption{\label{fig:DM}\footnotesize Lines showing the $\beta^\prime$ and $M_c$ values required to produce the correct relic abundance from PBH evaporation only. The DM particle we consider is a fermion with a mass of $1$ GeV. We show a range of PBH distributions, on the left we show log-normal distributions with a range of widths, $\sigma = 0.1, 0.5, 1.0, 2.0$ in orange, green, red and purple respectively. On the right we have a power law distribution with $\alpha=2.5$ and $\sigma=3$ in orange, the critical collapse (CC) distribution in green and the metric preheating (MP) distribution in red. In both panels the monochromatic PBH distribution is depicted with a blue dotted line for reference. The black dashed lines that overlay the solid lines show where warm DM constraints are in conflict with the PBH-produced DM. }
\end{figure*}
After the complete evaporation of the PBH population, we obtain the DM relic abundance in the standard manner. 
We present in Fig.~\ref{fig:DM} the PBH parameters which produce the observed relic abundance assuming a scalar DM particle with a mass of $1$ GeV for a log-normal distribution (left) and power law, critical collapse and metric preheating (MP) scenarios (right) while considering a monochromatic spin distribution with $a_\star = 0$.
For comparison purposes, we have included in both panels the results from a purely monochromatic distribution.
In the case where the mass distribution is a log-normal, we observe a crucial dependence on the width of such distributions.
For values $\sigma \lesssim 0.1$ we find that the initial PBH energy density needed to produce the correct abundance is quite similar to the monochromatic case.
Meanwhile, if the distributions are wider, that is, for $\sigma > 1.0$, the initial PBH fraction is significantly modified.
Such modifications occur because wider distributions contain a population of much heavier PBHs than the peak value $M_c$, these will produce much more DM particles because the number of emitted particles grows $\propto (\MBHi)^2$ in the case where DM mass is smaller than the initial PBH temperature $m_{\rm DM}<\TBHi$, cf.~Ref.~\cite{Cheek:2021odj}.
This implies that, for initial PBH densities which do not lead to a PBH domination, the $\beta^\prime$ value that leads to the observed DM abundance is reduced by a factor of $\sim{\cal O}(10^4)$ for $\sigma=2$ {in comparison with a Monochromatic mass distribution centered at $M=M_c$}.
Furthermore, the place where the PBH domination occurs, which leads to vertical relic abundance contours, is shifted to lower values of $M_c$ as the $\sigma$ becomes larger, due to the presence of heavier PBHs.
On the right panel of Fig.~\ref{fig:DM} we present the relic abundance contours for a power law distribution with $\alpha = 2.0, \sigma=2$ (orange), critical collapse (green) and metric preheating (red).
In the case of a power law mass distribution, we observe a similar effect as in the log-normal case: the presence of heavier PBHs enhances the DM production, thus requiring a smaller initial PBH fraction. 
In contrast, for the critical collapse and metric preheating cases, the relic contours do not differ in a noticeable way from the monochromatic case.
Since such distributions extend to lower PBH initial masses instead of higher ones, the DM production is dominated by the PBH having a mass $M_c$, corresponding to the peak of the distribution.
Furthermore, in both panels of Fig.~\ref{fig:DM} we overlay thick black dashed lines when the PBH produced DM is no longer consistent with observation and is too warm to support the small-scale structures observed in our Universe.

Of course, when DM particles emerge from the black holes, they are highly boosted. Through cosmic redshift these particles cool, and depending on their masses, can constitute warm ($\sim 10^{-2} -10^{3}\,{\rm GeV}$) or cold DM ($\gtrsim 10^{3}\,{\rm GeV}$)~\cite{Bode:2000gq,Baldes:2020nuv,Auffinger:2020afu, Masina:2021zpu, Cheek:2022dbx}. Previously, the mass constraints on $m_{\rm DM}$ have been applied for monochromatic distributions in mass and spin of the PBH. Since now our calculations allow for non-monochromatic distributions, we have adapted our interpretation of the warm DM constraints accordingly. Once again we make use of the computational tool {\tt CLASS}~\cite{CLASSI, CLASSII, CLASSIV} to determine the matter power spectrum in the CMB, where the input required for the DM phase space distribution, $f_{\rm DM}$, is
\begin{equation}
 f_{\rm DM}= \left.\frac{n_{\rm BH}\left(t_{\rm in}\right)}{g_{\rm DM}}\left(\frac{a(t_{\rm in})}{a(t)}\right)^3
\frac{1}{p^2}\frac{ d\mathcal{N}_{\rm DM}}{d p}\right\vert_{t=t_{\rm ev}},
\label{eq:DMphasespace}
\end{equation}
where $g_{\rm DM}$ is the number of degrees of freedom, $p$ is the three-momentum, and $n_{\rm DM}$ is the DM number density. With a PBH distribution, the particle emission rate per momentum is given by  
\begin{align}
 \frac{ d\mathcal{N}_{\rm DM}}{d p}&= \int d t^\prime\frac{a(t_{\rm fn})}{a(t')}\int d\asi\int {\rm d }\MBHi \nonumber\\
 &\times \mathcal F_{\rm in}\frac{ d^2\mathcal{N}_{\rm DM}}{d p^\prime  d  t^{\prime}}\left(p\frac{a(t_{\rm fn})}{a(t^\prime)},\MBH, \as  \right)\,,
 \label{eq:dNdP_redshift}
\end{align}
where similarly to above ${\rm d}^2\mathcal{N}_{\rm DM}/{\rm d}p^\prime {\rm d } t^{\prime}$ is a time-dependent function of $\MBH=\MBH(t,\MBHi, \asi,t_{\rm in})$ and $\as=\as(t,\MBHi, \asi,t_{\rm in})$. The $t_{\rm fn}$ here refers to the latest evaporation time we consider for a given distribution as described in section \ref{sec:numerical}. In practice we determine $f_{\rm DM}$ by running {\tt FRISBHEE} to evaluate $\mathcal F_{\rm in}$ and $a(t)$, which is then used to integrate numerically Eq.(\ref{eq:dNdP_redshift}). Also, the time-independent NCDM temperature is needed to interface the resulting  DM distribution with {\tt CLASS},
\begin{equation}
 \mathcal{T}_{\rm ncdm}=T_{\rm in}\frac{a(t_{\rm fn})}{T(t_0)}=\frac{T_{\rm in}}{T_{\rm ev}}\left(\frac{g_{s\star}(T_0)}{g_{s\star}(T_{\rm ev})}\right)^{1/3}\,,
 \label{eq:tncdm}
\end{equation}
where $T_{\rm ev}$ and $T_{\rm in}$ are the SM plasma temperatures at evaporation and PBH production respectively. Of course, because we are now working with distriutions of PBHs the evaporation temperature takes multiple values, $T_{\rm ev}$ here is simply the SM plasma temperature when the entire distribution has been evaporated, $T(t_{\rm fn})$. 

To determine whether a specific DM distribution is at odds with observations of structure in the Universe, such as those of the Lyman-$\alpha$ forest~\cite{Viel:2013fqw,Palanque-Delabrouille:2019iyz}, we use {\tt CLASS} to calculate the matter power spectrum $P(k)$,  quantifying the deviation from CDM by way of the transfer function, $T(k)$, defined in
\begin{equation}
 P(k)=P_{\rm CDM}(k)T^2(k)\,,
 \label{eq:transfer_func}
\end{equation}
where $k$ is the wavenumber. The scales at which $T(k)$ can start to stray from $1$ has been determined by the parameter fit 
\begin{equation}
 T(k)=(1+(\alpha k)^{2\mu})^{-5/\mu}\,,
 \label{eq:Transfer_param}
\end{equation}
where $\mu$ is dimensionless exponent which is fixed to $\mu=1.12$ as in Ref.~\cite{Viel:2005qj} and $\alpha$ is the breaking scale, which we take to be saturated at $\alpha=1.3\times 10^{-2}\,{\rm Mpc}\,h^{-1}$~\cite{Viel:2005qj, Palanque-Delabrouille:2019iyz, Garzilli:2019qki, Baldes:2020nuv}.\\

Returning to the results shown in Fig. ~\ref{fig:DM} we see that the warm DM constraints tend to be more constraining for wider distributions such as log-normal with $\sigma\gtrsim 1$ and a power law. This is because the heavier PBHs evaporate later, providing less time for DM to redshift, and it is exactly these heavier PBHs that produce the lion's share of DM. For the examples which reproduce a similar result for the monochromatic, the WDM constraints are almost identical. We would like to note, that taking the 1D fit for Eq.(\ref{eq:transfer_func}) may not be the most appropriate since when we take the 2D fit, we see some variation in the $\mu$ parameter, away from the quoted $\mu=1.12$. We leave a more sophisticated analysis for future work, but since the $\alpha$ parameter really controls where the matter power spectrum diverges from CDM, we are confident that the results we show here are correct to a reasonable degree of precision.

Due to the computational expense of performing the 3D integral, we have opted to only perform the above analysis in the Schwarzchild case, and we refer readers to Refs~\cite{Lennon:2017tqq, Masina:2021zpu, Cheek:2022dbx} to get an impression of how spin $\asi$, $\beta^\prime$ and $M_c$ parameters will affect the relic lines. We expect that Kerr distributions of BHs only have an appreciable effect on the relic lines for spin-2 DM. 

The main takeaway of this section is that broader distributions may well enable smaller $\beta^\prime$ values to produce the correct relic density, but this itself will introduce more aggressive warm DM constraints. 

For DM having a mass larger than the initial PBH temperature, we find that the overall behaviour is similar to the monochromatic scenario, i.e., depending on the DM mass, the relic density contours are either directly or inversely proportional to $M_c$; the change of behaviour occurs when $\TBHi=m_{\rm DM}$ and it is abrupt~\cite{Cheek:2021odj}. 
Such change of behaviour is much more gradual for wider distributions.

\begin{figure*}[t!]
    \centering
    \includegraphics[width=.475\linewidth]{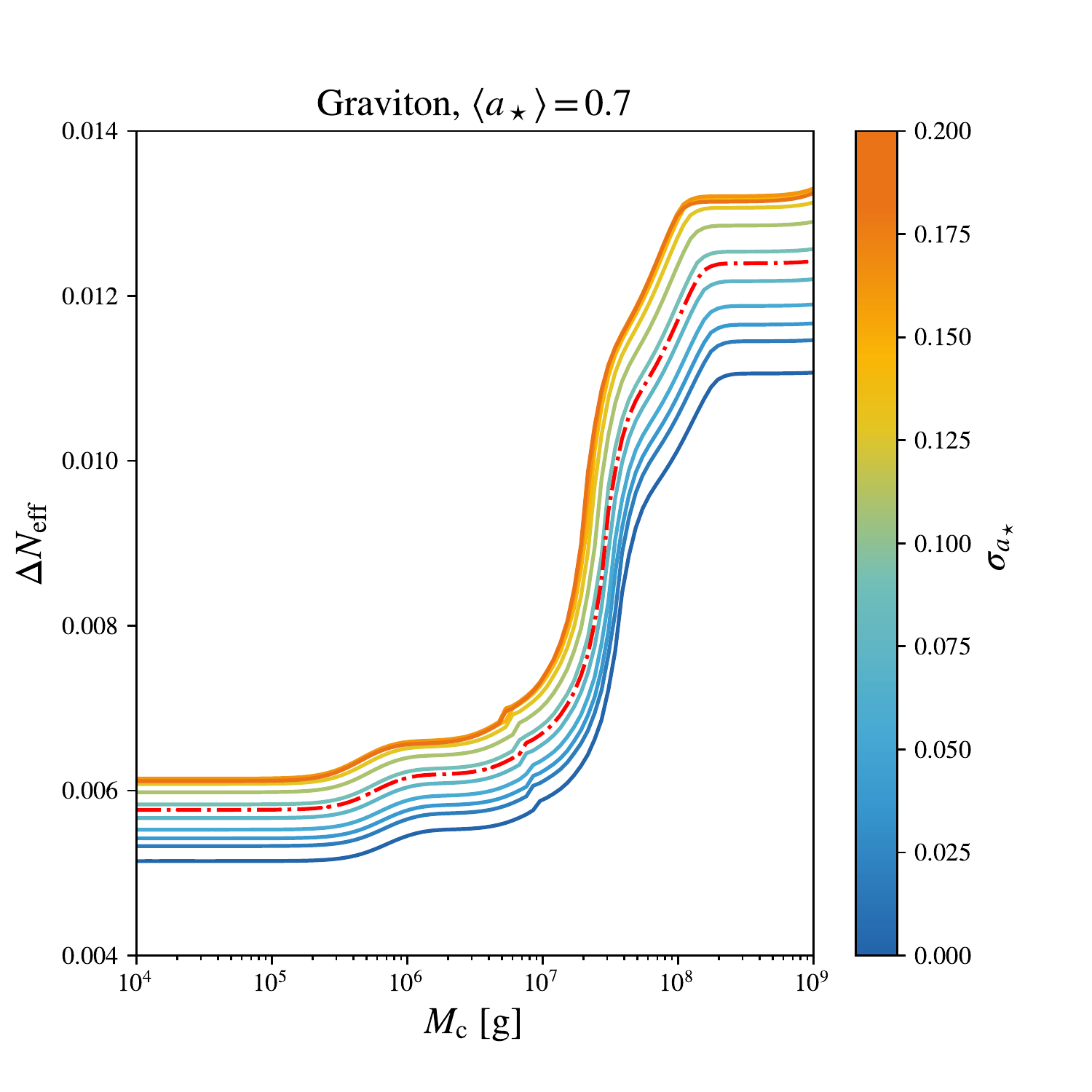}
    \includegraphics[width=.475\linewidth]{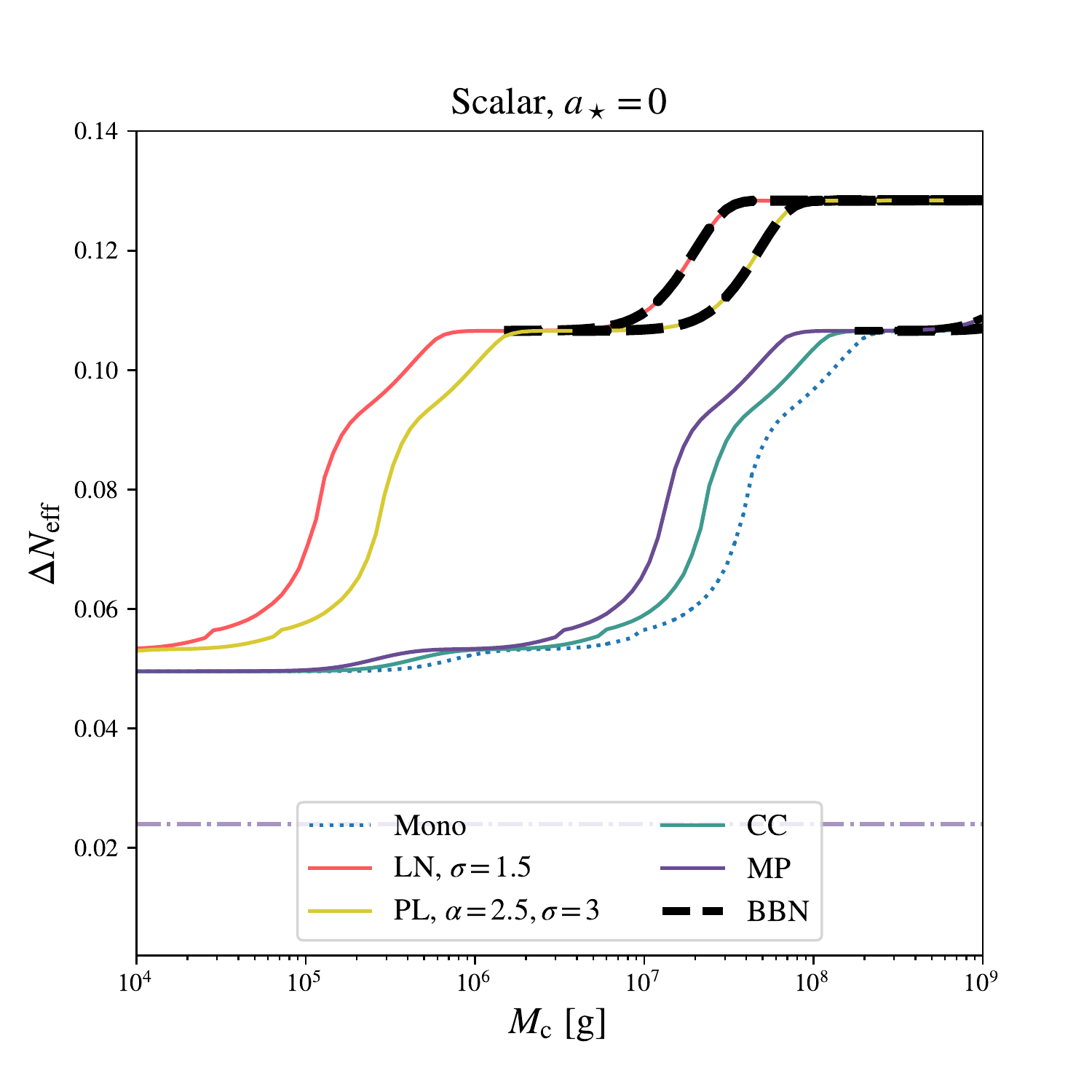}
    \caption{\label{fig:DR}\footnotesize Contribution to $\Delta N_{\rm eff}$ for gravitons (left) and new light scalar degrees of freedom (right) from the evaporation of PBHs. In the left panel, the mass distribution is log-normal, with a constant width $\sigma = 0.005$ and central mass $M_c$, and the colorbar represents the width of the spin distribution, taken to be gaussian, centered at $\langle a_\star\rangle = 0.7$, and with width $\sigma_{a_\star}$. The red dot-dashed line indicates the result for the merger spin distribution from Ref.~\cite{Fishbach:2017dwv}. On the right panel, we consider Schwarzschild PBHs with different examples of mass distributions, Log-normal (red), power-law (yellow), Critical Collapse (emerald), and Metric Preheating (purple). The black dashed lines indicate the values which are excluded from BBN limits.}
\end{figure*}

\subsection{Dark Radiation}
\label{sec:DR}

Similar to our previous discussion on DM generation, in this section we consider the emission of massless states that will modify the number of relativistic neutrino species, $\Neff$. 
To do so, we solve an  equation for the Dark Radiation (DR) comoving energy density
\begin{align}
    \dot{\varrho}_{\rm DR} &= - a(t)\int\int\frac{\varepsilon_{\rm DR}}{\varepsilon}\frac{d\MBH}{dt} {\cal F}_{\rm in}\,d\MBHi\,d\asi\,,
\end{align}
where now $\varepsilon_{\rm DR}$ corresponds to the contribution to the total evaporation function coming from the DR.
We track the evolution of all Universe species, and, after the full evaporation of the PBH population occurring when the plasma temperature is $T_{\rm ev}$, we determine the modification to $\Neff$, $\Delta \Neff$ as
\cite{Hooper:2019gtx}
\begin{widetext}
\begin{align}\label{eq:Neffg}
 \Delta N_{\rm eff} = \left\{\frac{8}{7}\left(\frac{4}{11}\right)^{-\frac{4}{3}}+N_{\rm eff}^{\rm SM}\right\} 
 \frac{\rho_{\rm DR}(T_{\rm ev})}{\rho_{\rm R}^{\rm SM}(T_{\rm ev})}
 \left(\frac{g_*(T_{\rm ev})}{g_*(T_{\rm eq})}\right)
 \left(\frac{g_{*S}(T_{\rm eq})}{g_{*S}(T_{\rm ev})}\right)^{\frac{4}{3}}\,,
\end{align}
\end{widetext}
where $N_{\rm eff}^{\rm SM} = 3.045$ are effective number of neutrinos~\cite{deSalas:2016ztq}, and $T_{\rm eq}=0.75$ eV is the matter-radiation equality temperature.

Let us focus first on the specific case where the DR particle is the spin-2 massless graviton, potentially the most well-motivated undiscovered particle in fundamental physics.
In Fig.~\ref{fig:DR} left, we present the modification on $\Neff$ coming from hot gravitons produced from the evaporation.
Here, we assumed a close-to-monochromatic mass distribution, $\sigma = 0.005$, and considered the scenarios where the spin distribution is gaussian (colored full lines) and for the case of the fourth generation merger distribution (red dashed line), taken from Ref.~\cite{Fishbach:2017dwv}.
For the gaussian case, the color indicates the value of the $\sigma_{a_\star}\in [0.0,0.2]$, where the value of zero corresponds to the monochromatic in spin situation.
The black dashed line corresponds to the monochromatic Schwarzschild case, included here for comparison.
We observe an enhancement on $\Delta \Neff$ for wider gaussian distributions since such wider distributions can increase the amount of emitted DR in comparison to the monochromatic case, depending on where the mean value lies. 
For instance, a Gaussian distribution with $\sigma_{a_\star}=0.1$, and mean value $\langle a_\star \rangle = 0.71$ will generate $\sim 34\%$ more gravitons than a monochromatic distribution with the same central value.
This leads to an increase of $\sim 18\%$ in $\Delta \Neff $ for $\sigma_{a_\star}=0.2$. 
Interestingly, for broader distributions, the contribution to $\Delta\Neff$ is reduced since those start to include PBHs which have a smaller angular momentum, for which the emission of gravitons is significantly decreased.
For other DR particles with lower spins, this dependence is less prominent.
We have found that for vectors, the enhancement only reaches $\sim 1\%$ for gaussian spin distributions with $\sigma_{s_\star} = 0.2$; for fermions and scalars the modification is smaller.
Such behavior is due to the mild dependence that the greybody factors have on the angular momentum parameter $a_\star$ for particles having lower spins~\cite{Page:1976ki}.

If we consider extended mass distributions instead, we observe a shift in the shape of the contribution to $\Delta \Neff$, see  the right panel of Fig.~\ref{fig:DR}.
The black dashed lines in the same figure indicate the parameters excluded by BBN constraints. 
We obtain such a limit by determining whether such PBH populations reheat the Universe after evaporation with a temperature $T_{\rm ev}$ smaller than the lower BBN bound of $\sim 5$~MeV~\cite{Hasegawa:2019jsa,Kawasaki:1999na,Kawasaki:2000en,Carr:2020gox}
Since the overall population would evaporate much later than in the monochromatic case, especially for broad distributions, the value of the evaporation temperature,  $T_{\rm ev}$, is reduced.
Therefore, the contribution from the dark radiation becomes more substantial.
Assuming a log-normal distribution, we find that the contribution to $\Delta \Neff$ is shifted by a factor of $\sim {\cal O}(10^3)$ to lower masses for $\sigma=1.5$.
For other types of mass distributions, such as the power-law, the behavior is similar.
Moreover, for distributions such as critical collapse and metric preheating (MP) the shift is less sizable since the evaporation temperature of such distributions is closer to the monochromatic one.


\subsection{Gravitational Waves from Evaporation}\label{sec:GWs}

There are several reasons why the presence of PBHs in the early Universe is expected to be accompanied by a sizeable spectrum of gravitational waves (GWs). First, it is clear that the production of PBHs, if it arises from the collapse of primordial perturbations, has to be accompanied by a large scalar power spectrum at a given scale. Such scalar perturbations are known to induce GWs at second order in perturbation theory~\cite{Baumann:2007zm, Saito:2008jc,Domenech:2021ztg,Kohri:2018awv}. However, the specific shape of the corresponding spectrum is entirely dependent on the shape of power spectrum considered, and therefore, it is not a unique prediction that can be obtained given a particular PBH distribution. Another interesting manner via which PBHs can produce gravitational waves is when they dominate the energy density of the Universe before they evaporate. In the latter case, the sudden transition from a matter to a radiation-dominated Universe, as the PBHs evaporate rapidly at the end of their lives,
can cause the gravitational potential  to oscillate (the so-called {\em poltergeist} mechanism) ~\cite{Inomata:2020lmk,Papanikolaou:2020qtd,Domenech:2020ssp,Domenech:2021wkk,Pearce:2022ovj,Bhaumik:2022pil, Papanikolaou:2022chm}. In this context, the faster the evaporation takes place, the sharper the transition between matter domination and radiation is, and the higher the peak of GWs expected to be probed by future observatories is. Therefore, it is expected that the extension of the mass and spin distributions, which mainly smears out the evaporation process (as we have seen in Sec.\ref{sec:cosmo}), has the tendency to suppress the GW spectrum. This can already be seen in Ref.~\cite{Inomata:2020lmk} --see also Ref.~\cite{Domenech:2021wkk}-- where it was shown that a width of order $\sigma = 10^{-2}$ for a log-normal distribution leads to a suppression of the GW spectrum of over three orders of magnitude. Another interesting aspect of the PBH formation and subsequent domination is that it can act as a source of isocurvature perturbation in the early Universe which also leads to a peak in the GW spectrum, at a different frequency than the effect described previously~\cite{Papanikolaou:2020qtd,Bhaumik:2022pil,Bhaumik:2022zdd}. Similarly, we expect that such a peak would be suppressed in the presence of a broader distribution of PBHs. Indeed, PBH of different masses are expected to form at different times, and the isocurvature perturbations that would be sourced by a smeared formation process are likely to lead to a broadened GW spectrum. For a fixed energy fraction of PBHs, this broadening of the spectrum has thus to be accompanied by a suppression of its overall amplitude. Such claim should in principle be verified numerically. However the study of induced gravitational waves from isocurvature perturbation requires a dedicated study that we plan to explore in a future work.

The last source of GWs arising from the evaporation of PBHs comes from the direct production of gravitons via Hawking evaporation~\cite{Dolgov:2011cq,Dong:2015yjs,Perez-Gonzalez:2020vnz}. The GW signal for a monochromatic mass spectrum with $\MBHi\sim 1\, (10^{4})\,\text{g}$ is expected in this case to peak at frequencies of order $10^{13}$ ($10^{15}$) Hz. While there are currently no technologies that can detect such high-frequency GW, there are proposed detectors that could, in principle, detect THz GWs \cite{Arvanitaki:2012cn,Ito:2019wcb,Chen:2020ler,Chou:2015sle,Page:2020zbr, Ghoshal:2022kqp}. 

\begin{figure*}[tb]
    \centering
    \includegraphics[width=.475\linewidth]{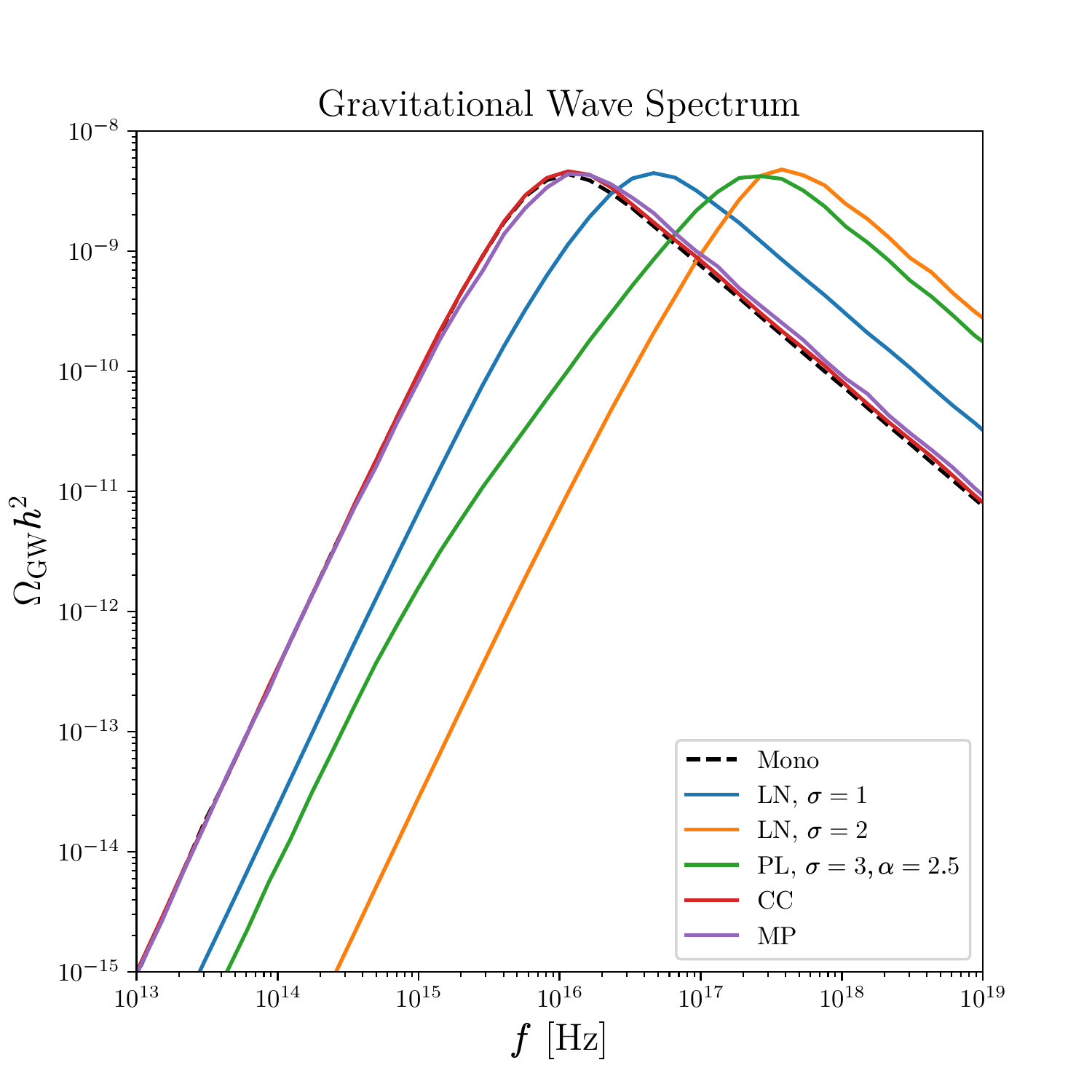}
    \includegraphics[width=.475\linewidth]{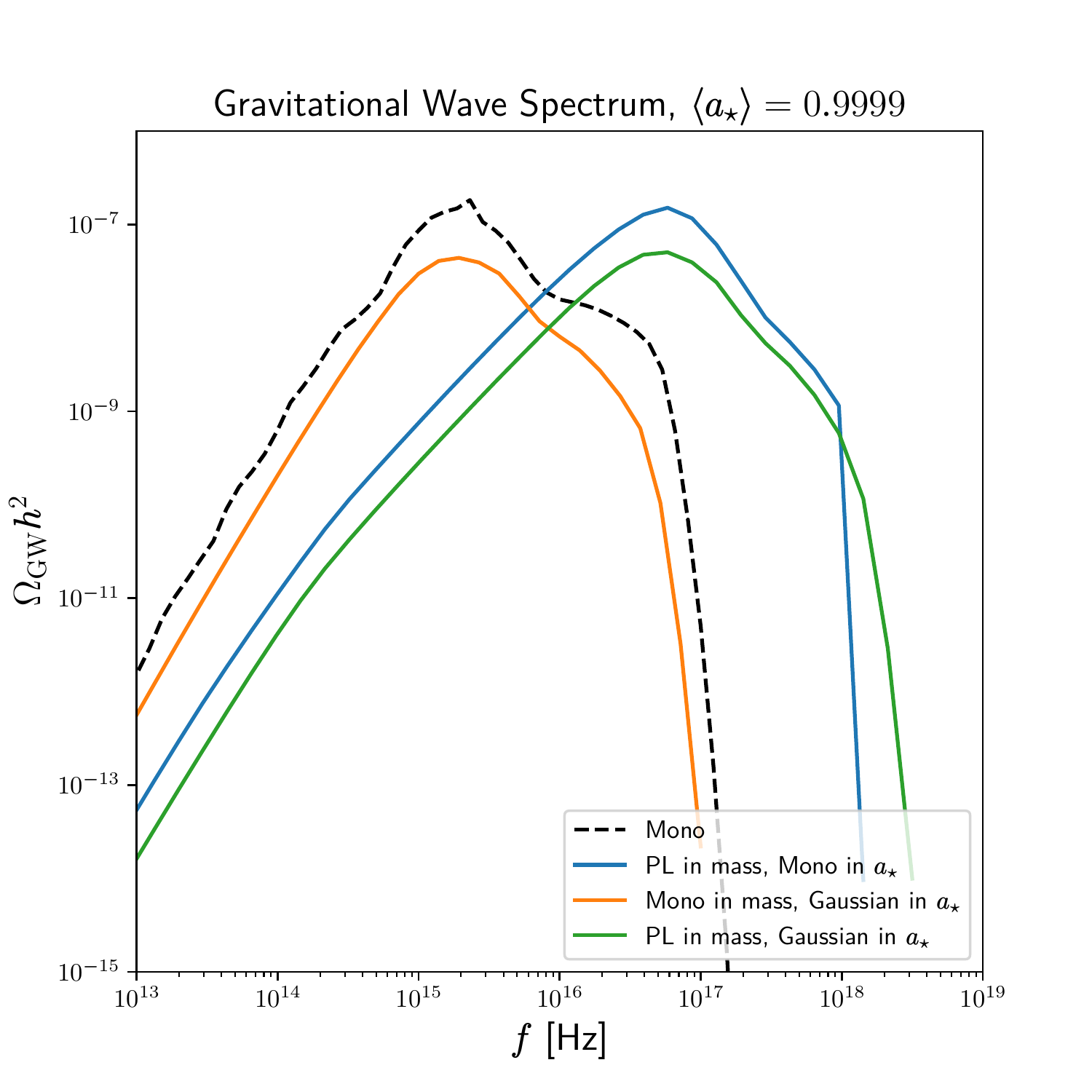}
    \caption{\label{fig:GWs}\footnotesize Gravitaional wave spectrum obtained from the evaporation of gravitons. We consider a variety of mass distributions (left),  $M_c=10^{4}\mathrm{g}$, and an initial PBH energy fraction of $\beta^\prime=10^{-7}$. We also exhibit the effect of spin for BH distributions (right), where the $M_c$ and $\beta^\prime$ is the same but $\langle a_\star\rangle=0.9999$. The Power-Law distribution in mass has parameters $\sigma=3$ and $\alpha=2.5$ and the Gaussian in $\as$ has the width $\sigma=0.1$.}
\end{figure*}
In the left panel of Fig.~\ref{fig:GWs}, we study the effect of having an extended PBH mass distribution on these high-frequency GWs. We use $M_c=10^{4}\mathrm{g}$, and an initial PBH energy fraction of $\beta^\prime=10^{-7}$, which are also parameters leading to a poltergeist signal detectable by LISA. We find that the GW spectrum due to direct Hawking radiation of gravitons, is not strongly affected by a non-monochromatic PBH distribution. The general trend is that for wider PBH distributions, the peak of the stochastic gravitational wave background (SGWB) shifts to either lower or higher frequencies, depending on whether the PBH population contains more light or heavy PBHs than the monochromatic case; moreover, the amplitude of the SGWB is not affected. Of the various observables we have studied, the SGWB produced from Hawking radiation is thus the least sensitive to the population of PBHs having a non-monochromatic spectrum, at least in terms of amplitude. This is interesting as it constitutes a robust signature of the existence of PBHs evaporating in the early Universe, as opposed to the other aforementioned GW signals. Furthermore, the fact that the peak frequency can change by orders of magnitude is important because it has implications on the physics reach of proposed THz GW detectors, which will have different frequency regions where they are optimally sensitive. 

For the case of GWs emitted by Kerr PBHs, Ref.~\cite{Dong:2015yjs} first computed the spectrum showing the modification coming from the enhancement due to the BH spin. We have reproduced their results for monochromatic mass and spin distributions. In the right panel of Fig.~\ref{fig:GWs}, we show the effect a population of highly spinning PBHs has on the SGWB signal. In order to exhibit how all effects intersect with each other we show the signals from monochromatic (black dashed), Power-Law distributed in $\MBH$ but monochromatic in $a_\star$ (blue), monochromatic in $\MBH$ but Gaussian distributed in $a_\star$ (orange), and Power-Law in $\MBH$ and Gaussian in $\as$ (green).   As was shown in Ref.~\cite{Dong:2015yjs}, the amplitude of the SGWB is enhanced substantially when $a_\star$ is close to maximal, comparing the left panel to the right, we observe that the peak amplitude is $\sim10^2$ larger. The peak frequency is at a similar position, but generally, the spectral shape of the SGWB is quite different thanks to the nature of the graviton emission spectrum as such high $\as$ values. We see that introducing a Gaussian in distributed $\as$ weakens the strength of the SGWB signal because fewer PBHs are maximally spinning. When this effect is in concert with mass distributed PBH population, we can see that the effect is similar to the Schwarzschild case, whereby the mass distribution causes the SGWB signal to shift in frequency but the amplitude remains unchanged.

\section{Summary and Conclusion}
\label{sec:conc}
In this paper, we have explored the effect of PBH mass and spin distributions on various cosmological observables. We outlined our numerical method, which simultaneously evolves the mass and spin distributed PBH population in time and focused on well-studied mass  (log-normal, power law, critical collapse and metric preheating) and spin distributions (spin distribution from mergers and Gaussian). 
While the results are unsurprising, to our knowledge, this work is the first attempt at numerically solving the evaporation of both the mass and spin of PBHs together with the Friedmann-Boltzmann equations that describe the time evolution of different Universe's components.
We found that non-monochromatic mass and spin distributions reduced how suddenly the transition from PBH to radiation domination occurred. Naturally, the wider the distribution the more slowly such a transition occurs. In the case of light fermionic DM production from PBHs, we studied Schwarzschild PBHs with a non-trivial mass distribution. We found that broader mass distributions required a smaller initial number density to produce the observed relic density, as compared to a monochromatic distribution centered at $M=M_c$. Intuitively, this occurs as a broader mass distribution includes heavier PBHs, and thus requires less to produce the same overall energy density in PBHs. Likewise, we found that the  warm DM constraint was significantly more stringent for broad distributions because the heavier PBHs would evaporate later and thus  diminishing the effect of redshifting the boosted DM. We studied the effect of a non-trivial mass and spin distribution on $\Delta N_{\rm eff}$. We   fixed the mass distribution and allowed the width of the spin distribution to vary  and vice versa. Widening the spin distribution tends to produce an increase in $\Delta N_{\rm eff}$ as there are a higher proportion of higher spin PBHs which are efficient at producing gravitons. The effect of increasing the spin width , $\sigma_{a_{\star}}$, from $0$ to $0.2$ has the effect of increasing the contribution to $\Delta N_{\rm eff}$ by $\sim 10\%$. On the other hand, for a fixed central mass of PBH, for a Schwarzchild PBH, having a non-trivial mass distribution can increase the contribution to $\Delta N_{\rm eff}$ by a factor of a few (see the right panel of  \figref{fig:DR}). Finally, we consider the effect finite mass and spin PBH widths had on the SGWB produced from direct Hawking radiation of the PBHs. 
We found that, unlike for other GW signals predicted from the evaporation of PBHs, the amplitude of the graviton production via Hawking emission is mainly insensitive to the extension of the PBH mass distribution. This suggests that the production of GWs from the evaporation of PBHs is amongst the most robust signals that could be searched for with observations in the future regarding the existence of light PBHs in the early Universe. Our code {\tt FRISBHEE} \href{https://github.com/yfperezg/frisbhee}{\faGithub} is publicly available at the address \href{https://github.com/yfperezg/frisbhee}{https://github.com/yfperezg/frisbhee}.

\section*{Acknowledgments}
AC is supported by the grant ``AstroCeNT: Particle Astrophysics Science and Technology Centre" carried out within the International Research Agendas programme of the Foundation for Polish Science financed by the European Union under the European Regional Development Fund. 
The work of LH and YFPG is funded by the UK Science and Technology Facilities Council (STFC) under grant ST/P001246/1. LH acknowledges the support of the Institut Pascal at Université Paris-Saclay during the Paris-Saclay Astroparticle Symposium 2022, with the support of the IN2P3 master projet UCMN, the P2IO Laboratory of Excellence (program “Investissements d’avenir” ANR-11-IDEX-0003-01 Paris-Saclay and ANR-10-LABX-0038), the P2I axis of the Graduate School Physics of Université Paris-Saclay, as well as IJCLab, CEA, IPhT, APPEC,  and ANR-11-IDEX-0003-01 Paris-Saclay and ANR-10-LABX-0038.
This project has received funding/support from the European Union’s Horizon 2020 research and innovation programme under the Marie Sk\l{}odowska-Curie grant agreement No 860881-HIDDeN.
This work used the DiRAC@Durham facility managed by the Institute for Computational Cosmology on behalf of the STFC DiRAC HPC Facility (\href{www.dirac.ac.uk}{www.dirac.ac.uk}). The equipment was funded by BEIS capital funding via STFC capital grants ST/P002293/1, ST/R002371/1 and ST/S002502/1; Durham University and STFC operations grant ST/R000832/1. DiRAC is part of the National e-Infrastructure. 
This work has made use of the Hamilton HPC Service of Durham University.

\appendix

\section{Mass and spin evolution}\label{ap:KPBH_sol}

In the limit of relativistic evaporation products (with masses $m\ll T_{\rm BH}$), the system of equations presented in Eqs.~\eqref{eq:dynamicsKerr} takes the simple form
\bea\label{eq:2Dsystem}
\frac{d \MBH}{d t}&=&-\epsilon(a_\star)\frac{M_p^4}{\MBH^2}\,.\\
\frac{d a_\star}{d t}&=&-a_\star\frac{Mp^4}{\MBH^3}\left[\gamma(a_\star)-2\epsilon(a_\star)\right]\
\eea
where $\epsilon(a_\star)\equiv \epsilon(M_{\rm BH},a_\star)$ and $\gamma(a_\star)\equiv \gamma(M_{\rm BH},a_\star)$ are now independent of $\MBH$, which can be used to simplify greatly the study of the system's dynamic~\cite{Page:1976ki}. Before studying the time evolution of the distribution $\fBH(\MBH,a_\star,t)$, let us first recall how to simply obtain  the time evolution of any pair $(\MBH(t),a_\star(t))$ starting with initial condition $\left(\MBHi, a_\star^i\right)$ at time $t_{\rm in}$, using the formalism introduced in \cite{Page:1976ki}. 
First of all, the spin can be used as a new time coordinate by defining
\be
y\equiv-\ln(\as)\,.
\ee
Then, let us consider the generic solution starting at time $t=0$ with $(\MBH,\as)=({\MBH}_1,1)$. After defining the mass ratio function $z$ as
\be
z\equiv - \ln\left(\frac{\MBH}{{\MBH}_1}\right)\,,
\ee
and the time as
\be
\tau \equiv M_1^{-3} t\,,
\ee
one can search for the solution of the system of Eq.~\eqref{eq:2Dsystem} now re-written as
\bea\label{eq:diffsys}
\frac{d z}{d y} &=& \frac{\epsilon(\as)}{\gamma(\as)-2\epsilon(\as)}\,,\nonumber\\
\frac{d \tau}{d y} &=& \left(\frac{\MBH}{{\MBH}_1}\right)^3\frac{1}{\gamma(\as)-2\epsilon(\as)}\,,
\eea
using the initial conditions $\tau = z = 0$ at $\ y\to -\infty$. Once this solution is found, it is remarkable that any solution $(\MBH,\as)$ starting with initial conditions $(\MBHi,\asi)$ at time $t=t_{\rm in}$ and satisfying Eq.~\eqref{eq:2Dsystem} can be obtained by simply computing
\bea\label{eq:evo}
\MBH &=& \MBHi e^{z_{\rm in}-z}\,,\nonumber\\
(t-t_{\rm in})&=&(\MBHi)^3e^{3z_{\rm in}}(\tau-\tau_{\rm in})\,,
\eea
where $z$ is the unique solution of the system \eqref{eq:diffsys} satisfying $\tau = z = 0$ at $\ y\to -\infty$ and $z_{\rm in}$ and $y_i$ are defined as
\bea
z_{\rm in}&\equiv& z\left[-\ln(\asi)\right]\,,\\
\tau_{\rm in}&\equiv& \tau\left[-\ln(\asi)\right]\,.
\eea

In order to trace the evolution of the distribution function $\fBH(\MBH, \as,t)$ one is to compute the time evolution of the surface element $d \MBH d \as$ and impose that the infinitesimal number of PBHs that sit in this elementary surface remains constant with time:
\bea\label{eq:condition}
a^3(t)d n_{\rm BH} &\equiv& a^3(t)\fBH(\MBH,\as,t)d \MBH d \as\nonumber\\
&=& a^3(t_{\rm in})\fBH(\MBHi, \asi,t_{\rm in})d \MBHi d \asi\,.\nonumber\\
\eea
At a fixed time, $t$, the set of equations of \equaref{eq:evo} provide implicit relations between $(\MBH,\as)$ and $(\MBHi,\asi)$, meaning that one can perform a time-dependent change of coordinate from one set of variables to the other. Using the Jacobian
\be
\mathcal J \equiv \left|\frac{\partial \MBH}{\partial \MBHi}\frac{\partial \as}{\partial \asi}-\frac{\partial \MBH}{\partial \asi}\frac{\partial \as}{\partial \MBHi}\right|\,.
\ee
one then simply obtain that
\be
d \MBH d \as = \mathcal J d \MBHi d \asi\,.
\ee
From Eq.~\eqref{eq:condition} we can thus deduce that
\be\label{eq:change}
a^3(t)\fBH(\MBH, \as, t) = a^3(t_{\rm in})\frac{\fBH(\MBHi,\asi, t_{\rm in})}{\mathcal J}\,.
\ee

\section{Calculation of the Jacobian} \label{ap:jaco}
Let us first recall Eq.~\eqref{eq:evo}:
\bea\label{eq:evo2}
\MBH &=& \MBHi e^{z_{\rm in}-z}\,,\nonumber\\
(t-t_{\rm in})&=&(\MBHi)^3e^{3z_{\rm in}}(\tau-\tau_{\rm in})\nonumber\,.
\eea
In this equation, $z$ and $\tau$ are functions of $a$, meaning that $z_{\rm in}$ and $\tau_{\rm in}$ are independent of $\MBHi$:
\bea
\pd{z_{\rm in}}{\MBHi}&=&\pd{\tau_{\rm in}}{\MBHi}=0\,,\nonumber\\
\pd{z}{\MBH}&=&\pd{\tau}{\MBH}=0\,.
\eea
Moreover, one can use the fact that $\tau$ and $z$ are solutions of the system \eqref{eq:diffsys} to write
\bea\label{eq:der1}
\pd{z}{\MBHi}&=&\pd{y}{\MBHi}\fd{z}{y}=-\frac{1}{\as}\pd{\as}{\MBHi}\frac{\epsilon}{\gamma-2\epsilon}\,,\nonumber\\
\pd{\tau}{\MBHi}&=&\pd{y}{\MBHi}\fd{\tau}{y}=-\frac{1}{\as}\pd{\as}{\MBHi}\!\!\left(\!\frac{\MBH}{\MBHi}\!\right)^3\!\!\!\!\frac{1}{\gamma-2\epsilon}\,,
\eea
and
\bea\label{eq:der2}
\pd{z}{\asi}&=&\pd{y}{\asi}\fd{z}{y}=-\frac{1}{\as}\pd{\as}{\asi}\frac{\epsilon}{\gamma-2\epsilon}\,,\nonumber\\
\pd{\tau}{\asi}&=&\pd{y}{\asi}\fd{\tau}{y}=-\frac{1}{a}\pd{a}{\asi}\!\left(\!\frac{\MBH}{\MBHi}\!\right)^3\!\!\!\!\frac{1}{\gamma-2\epsilon}\,.
\eea

Deriving the first line of Eq.~\eqref{eq:evo2} one can obtain the relations
\bea
\pd{\MBH}{\MBHi}&=&\frac{\MBH}{\MBHi}+\frac{\MBH}{\as}\pd{\as}{\MBHi}\frac{\epsilon}{\gamma-2\epsilon}\,,\\
\pd{\MBH}{\asi}&=&\left(-\frac{1}{\asi}\frac{\epsilon_{\rm in}}{\gamma_{\rm in}}+\frac{1}{\as}\pd{\as}{\asi}\frac{\epsilon}{\gamma-2\epsilon}\right)\MBH\,.
\eea
Similarly, the second line of Eq.~\eqref{eq:evo2} provides
\bea
\pd{\tau}{\MBHi}&=&-3(\MBHi)^{-4}(t-t_{\rm in})e^{-3 z_{\rm in}}\,,\nonumber\\
\pd{\tau}{\asi}&=&\frac{3}{\asi}\frac{\epsilon_{\rm in}}{\gamma_{\rm in}-2\epsilon_{\rm in}}(\MBHi)^{-3}(t-t_{\rm in})e^{-3 z_{\rm in}}\nonumber\\
&&-\frac{1}{\asi}\left(\frac{\MBHi}{\MBH}\right)^3\left.\frac{1}{\gamma-2\epsilon}\right|_{a=\asi}\,.
\eea
From those two expressions, together with Eq.~\eqref{eq:der1}-\eqref{eq:der2}, we can extract
\bea
\pd{\as}{\MBHi}
&=&\frac{3a(\gamma-2\epsilon)}{(\MBHi)^4}\left(\frac{\MBHi}{\MBH}\right)^3(t-t_{\rm in})\,,\nonumber\\
\pd{\as}{\asi} 
&=&\frac{\as}{\asi}\frac{\gamma-2\epsilon}{\gamma_{\rm in}-2\epsilon_{\rm in}}\left(\frac{1}{\MBH}\right)^3\left[-3\epsilon_{\rm in}(t-t_{\rm in})+\MBHi^3\right]\,.\nonumber\\
\eea

Using those expressions, one can compute
\bea
\mathcal J &\equiv&  \left|\frac{\partial \MBH}{\partial \MBHi}\frac{\partial \as}{\partial \asi}-\frac{\partial \MBH}{\partial \asi}\frac{\partial \as}{\partial \MBHi}\right|\,,\nonumber\\
&=&\left|\frac{\as}{\asi}\frac{\gamma-2\epsilon}{\gamma_{\rm in}-2\epsilon_{\rm in}}\left(\frac{\MBHi}{\MBH}\right)^2\right|\,,
\eea

\section{Number Density Conservation}\label{ap:Nconservation}
When performing an integral over $(\MBH,a)$ at time $t$, it is convenient to use the Jacobian $\mathcal J$ in order to change variables, and integrate over $(\MBHi, \asi)$ at time $t_{\rm in}$. To see how this change of variable operates, let us first write down the comoving number density of PBH as
\be
a(t)^3 n_{\rm BH}(t) \equiv a(t)^3\int_{0}^\infty\int_0^1\fBH(\MBH,\as,t)d \MBH d\as\,.
\ee
Using Eq.~\eqref{eq:change} we obtain
\bea
a(t)^3 n_{\rm BH}(t) &=& a(t_{\rm in})^3\int_{0}^\infty\int_0^1\frac{\fBH(\MBHi,\asi,t_{\rm in})}{\mathcal J}\mathcal J d \MBHi d \asi\,,\nonumber\\
&=& a(t_{\rm in})^3\int_{0}^\infty\int_0^1\fBH(\MBHi,\asi,t_{\rm in}) d \MBHi d \asi\,,\nonumber\\
&=& a(t_{\rm in})^3 n_{\rm BH}(t_{\rm in})\,.
\eea
As expected, the number of comoving PBHs is thus conserved. 

\vfill

\bibliographystyle{apsrev4-1}
\bibliography{main.bib}

\end{document}